\documentclass[onecolumn,a4paper,12pt,oneside]{article} 
\usepackage[top=2.5cm,left=2.5cm,right=2.5cm,bottom=3cm]{geometry}    
\usepackage{draftwatermark} 
\SetWatermarkText{\bf Preprint}
\SetWatermarkAngle{45}
\SetWatermarkScale{3}
\SetWatermarkLightness{0.85}     
\usepackage[parfill]{parskip}    
\usepackage{graphicx}
\usepackage{amssymb}
\usepackage{amsmath}
\usepackage{amsfonts}
\usepackage{amsthm}
\usepackage{epstopdf}
\usepackage[usenames,dvipsnames,svgnames,table]{xcolor}
\usepackage{tikz}
\usepackage{array}
\usepackage[T1]{fontenc}
\usepackage[utf8]{inputenc}
\usepackage{booktabs} 
\usepackage{array} 
\usepackage{paralist} 
\usepackage{verbatim} 
\usepackage{subfig} 
\usepackage[colorlinks=true]{hyperref}
\hypersetup{colorlinks=true, linktocpage=true, linkcolor=Blue, citecolor=Green,
hypertexnames=true, urlcolor=MidnightBlue,pdftex}
\usepackage[english]{babel}
\usepackage{wrapfig}
\usepackage{multirow}
\usepackage{quoting}
\usepackage{rotating}
\quotingsetup{font=normalsize}
\theoremstyle{plain}

\usepackage{bm}
\usepackage{abstract}
\usepackage{cite}
\usepackage{float}
\usepackage{textcomp} 	
\usepackage{tabularx} 
\usepackage{caption}
\usepackage{authblk} 
\usepackage{eso-pic} 
\usepackage{lineno} 
\providecommand{\keywords}[1]{\textbf{{Key words: }} #1} 

\newcommand{\be}{\begin{equation}}
\newcommand{\ee}{\end{equation}}
\newcommand{\bs}{\begin{split}}
\newcommand{\es}{\end{split}}

\newcommand{\beps}{\boldsymbol{\varepsilon}}
\newcommand{\balpha}{\boldsymbol{\alpha}}

\newcommand{\bk}{\boldsymbol{\kappa}}
\newcommand{\bgamma}{\boldsymbol{\gamma}}

\renewcommand{\Phi}{\varPhi}
\newcommand{\gr}{\mathbf}

\renewcommand{\Theta}{\varTheta}
\renewcommand{\Psi}{\varPsi}
\renewcommand{\Sigma}{\varSigma}
\newcommand{\A}{\mathbb{A}}
\newcommand{\Ac}{\mathcal{A}}
\newcommand{\B}{\mathbb{B}}
\newcommand{\Bc}{\mathcal{B}}

\newcommand{\D}{\mathbb{D}}
\newcommand{\Dc}{\mathcal{D}}
\newcommand{\Q}{\mathbb{Q}}

\renewcommand{\Delta}{\varDelta}
\renewcommand{\phi}{\varphi}
\renewcommand{\psi}{\varPsi}

\newcommand{\N}{\mathbf{N}}
\newcommand{\M}{\mathbf{M}}
\newcommand{\U}{\mathbf{U}}
\newcommand{\V}{\mathbf{V}}
\newcommand{\W}{\mathbf{W}}

\renewcommand{\u}{\mathbf{u}}
\renewcommand{\v}{\mathbf{v}}
\newcommand{\w}{\mathbf{w}}

\title{\LARGE{\bf Coupled thermoelastic isotropic laminates}}

\author{Paolo Vannucci\\ 
\begin{small}LMV - Laboratoire de Mathématiques de Versailles, UMR8100.\\Université de Versailles et Saint Quentin - \href{mailto:paolo.vannucci@uvsq.fr}{paolo.vannucci@uvsq.fr}\bigskip\bigskip\\
Final version in Journal of Engineering Mechanics, \href{https://doi.org/10.1061/JENMDT.EMENG-7593}{https://doi.org/10.1061/JENMDT.EMENG-7593}\end{small}}

\begin{document}
\maketitle

\hrule
\begin{abstract}
We consider in this paper the general properties of laminates designed to be isotropic in extension and in bending and with a coupling between the in- and out-of plane responses. In particular, we analyze the mathematical properties of the tensors describing the elastic and thermal behavior and the mechanical consequences of these properties. The differences, from the mathematical and mechanical point of view, between the hybrid laminates, i.e. composed by layers of different materials, and those made of identical plies, are pointed out and analyzed. The polar formalism for planar tensors is used in this study.

\end{abstract}
\medskip
\hrule
\bigskip

\section{Introduction}
Laminates are widely used in  applications, e.g. in aircraft construction, automotive, sport devices etc. In several cases, the objective is to obtain a stiff and  light structure, and isotropy can be the good choice whenever there is not a prevailing direction of the applied loads. That is why isotropic laminates have been studied in the past, see e.g. \cite{werren53,Fukunaga90a,Wu-Avery,Wu79,Paradies96,Grediac99b,CompStruct02}.

Like for the other types of laminates, isotropic laminated plates  are normally uncoupled, i.e. there is not coupling between the in- and out-of-plane responses. However, coupling can be an interesting property to obtain some special mechanical effects, and it deserves to be studied. Very few are the papers concerning coupling and most important, the exploration of the mathematical and mechanical aspects of coupling is in its debut. The relation between the stacking sequence and the amount of coupling has been first studied in \cite{vannucci12joe1}, where a parameter called {\it degree of coupling} was introduced. More recently, a set of papers concerning the effects of coupling and its mathematical aspects have been studied: in \cite{vannucci23}  the bounds of the coupling tensor have been studied for the first time, while in \cite{vannucci23a} the relations between the stiffness coupling tensor and the compliance one, and also with the extension and bending behaviors, and the connected mathematical aspects, have been examined. Finally, in  \cite{vannucci23b}, the same analysis have been conducted considering also the thermal aspects. 

In this paper, we extend the previous analyses to the particular case of isotropic laminates. In this study, unlike in the previous ones, we also consider the case of {\it hybrid laminates}, i.e. of plates obtained stacking together layers of different materials. This is done because in the particular case of isotropy, some applications of this type really exist, e.g.  bimetal plates, and also because the choice, once and for all, of the type of elastic symmetry, i.e. of isotropy, makes simpler the complete analysis of the problem. However, it is mostly in a case like this one that the same classical concept of material symmetry fails. If we accept to consider a laminate as a {\it meta-material}, or in other common jargons, a {\it complex material} or an {\it architectured material}, then it is just analyzing the different cases of isotropic laminates proposed in the paper that we can see how problematic is to define a correct class of symmetry for such a solid. Coupling interacts with extension and bending stiffnesses, giving, in general, different elastic symmetries for the stiffness and compliance tensors. Moreover, depending upon the type of laminate, i.e. if it is or not hybrid, and that of the layers, this  interaction changes.

In short, it is questionable how to correctly define the elastic symmetries of a coupled laminate. That is why in this paper we have taken a precise point of view: we define as isotropic a laminate whose stiffness tensors for extension and bending are both isotropic.   This nothing implies concerning the coupling tensor and all the compliance tensors. The situations can be very different and we can identify four principal types of isotropic laminates, that are
\begin{enumerate}[a)]
\item isotropic hybrid laminates composed of anisotropic plies;
\item isotropic hybrid laminates composed of isotropic plies;
\item bimetal plates;
\item isotropic laminates made of identical anisotropic plies.
\end{enumerate}
Before analyzing separately these four categories of laminates, we give below a rapid recall of the governing theories and of the polar formalism, used for this research.

\section{The governing equations}
In the classical theory,  the constitutive law of laminates is, \cite{jones,vannucci_libro},
 \begin{equation}
 \label{eq:fundlawtherm}
 \left\{\begin{array}{c}\gr{N} \\\hline \gr{M}\end{array}\right\}=
 \left[\begin{array}{c|c}h\A & \frac{h^2}{2}\B \\\hline \frac{h^2}{2}\B & \frac{h^3}{12}\D\end{array}\right] \left\{\begin{array}{c}\beps \\\hline \bk\end{array}\right\}
 -t \left\{\begin{array}{c}h\gr{U} \\\hline \frac{h^2}{2}\gr{V}\end{array}\right\}-\nabla t \left\{\begin{array}{c}\frac{h^2}{2}\gr{V} \\\hline \frac{h^3}{12}\gr{W}\end{array}\right\};
 \end{equation}
where:
\begin{itemize}
\item $\N$ and $\M$ are respectively the extension forces and bending moments  tensors;
\item $\beps$ and $\bk$ are respectively the in-plane deformation and the curvature tensors;
\item $h$ is the laminate's thickness;
\item $t$ is the temperature variation with respect to the temperature corresponding to a no strain state;
\item $\nabla t$ is the through-the-thickness gradient of temperature;
\item $\A,\B$ and $\D$ are, respectively, the extension, coupling and bending stiffness tensors;
\item $\U,\V$ and $\W$ are the analogous of $\A,\B$ and $\D$ respectively, but for a thermal action.
\end{itemize}

$\A,\B$ and $\D$ are elasticity tensors and as such they have the minor and major symmetries, i.e.,
\be
\A_{ijkl}=\A_{jikl}=\A_{ijlk},\ \A_{ijkl}=\A_{klij}.
\ee
The existence of the major symmetries is the condition for a 4th-rank tensor to be symmetric: $\A=\A^\top$, cf. \cite{vannucci_alg}. Of course,  the same is true  also for $\B$ and $\D$. 

 $\U$ determines the in-plane forces due to a through-the-thickness uniform change of temperature $t$, $\W$ the couples linked to a temperature gradient  $\nabla t$ and $\V$ is the thermoelastic coupling stiffness  tensor: it determines the in-plane forces due to a temperature gradient $\nabla t$ and the couples  caused by a uniform change of temperature $t$. All of them are  second-rank symmetric tensors.

The general expressions of $\A,\B,\D$ are 
 \begin{equation}
 \label{eq:compABD}
 \begin{split}
 &\A=\frac{1}{h}\sum_{k=1}^n\int_{z_{k-1}}^{z_k}\Q(\delta_k)dx_3=\frac{1}{h}\sum_{k=1}^n({z_k}-{z_{k-1}})\Q(\delta_k),\\
 &\B=\frac{2}{h^2}\sum_{k=1}^n\int_{z_{k-1}}^{z_k}x_3\Q(\delta_k)dx_3=\frac{1}{h^2}\sum_{k=1}^n({z_k^2}-{z_{k-1}^2})\Q(\delta_k),\\
& \D=\frac{12}{h^3}\sum_{k=1}^n\int_{z_{k-1}}^{z_k}x_3^2\Q(\delta_k)dx_3=\frac{4}{h^3}\sum_{k=1}^n({z_k^3}-{z_{k-1}^3})\Q(\delta_k),
 \end{split}
 \end{equation}
with $\mathbb{Q}(\delta_k)$ the reduced stiffness tensor of the $k-$th layer, at the orientation $\delta_k$, $n$ the number of plies  and $z_k,z_{k-1}$ the position of the upper and lower surfaces of the same layer on the thickness of the plate. 
Similarly,
\be
\label{eq:tensorthemoelast}
\begin{split}
&\gr{U}=\frac{1}{h}\sum_{k=1}^n\int_{z_{k-1}}^{z_k}\gamma_k(\delta_k)dx_3=\frac{1}{h}\sum_{k=1}^n(z_k-z_{k-1})\gamma_k(\delta_k),\\
&\gr{V}=\frac{2}{h^2}\sum_{k=1}^n\int_{z_{k-1}}^{z_k}x_3\gamma_k(\delta_k)dx_3=\frac{1}{h^2}\sum_{k=1}^n(z_k^2-z_{k-1}^2)\gamma_k(\delta_k),\\
&\gr{W}=\frac{12}{h^3}\sum_{k=1}^n\int_{z_{k-1}}^{z_k}x_3^2\gamma_k(\delta_k)dx_3=\frac{4}{h^3}\sum_{k=1}^n(z_k^3-z_{k-1}^3)\gamma_k(\delta_k),
\end{split}
\ee
where the tensor $\bgamma(\delta_k)$ is given by
 \be
\label{eq:tensorgammatermico}
\bgamma(\delta_k)=\mathbb{Q}(\delta_k)\balpha(\delta_k), \ \ k=1,...,n,
\ee
with $\balpha(\delta_k)$ the tensor of the thermal expansion coefficients of $k-$th layer, at the orientation $\delta_k$.

The converse of the constitutive equation (\ref{eq:fundlawtherm}) is, cf. \cite{vannucci_libro},
 \begin{equation}
  \label{eq:fundlawinversetherm}
 \left\{\begin{array}{c}{\beps} \\\hline \bk\end{array}\right\}=
 \left[\begin{array}{c|c}\frac{1}{h}\mathcal{A} & \frac{2}{h^2}\mathcal{B} \\\hline \frac{2}{h^2}\mathcal{B}^\top & \frac{12}{h^3}\mathcal{D}\end{array}\right]
 \left\{\begin{array}{c}\gr{N} \\\hline \gr{M}\end{array}\right\}+t \left\{\begin{array}{c} \gr{u} \\\hline \gr{v}_1\end{array}\right\}+\nabla t \left\{\begin{array}{c}\gr{v}_2 \\\hline \gr{w}\end{array}\right\},
 \end{equation}
with $\Ac,\Bc$ and $\Dc$  the  compliance corresponding of $\A,\B$ and $\D$, given in the order by
  \begin{equation}
 \label{eq:inversetensorsABD}
 \begin{split}
 &\mathcal{A}=(\A-3\B\D^{-1}\B)^{-1},\\ 
 &\mathcal{B}=-3\mathcal{A}\B\D^{-1}=(-3\mathcal{D}\B\A^{-1})^\top=-3\A^{-1}\B\mathcal{D},\\
  &\mathcal{D}=(\D-3\B\A^{-1}\B)^{-1},
\end{split}
 \end{equation}
 and
\be
\label{eq:tenstherminverses}
\begin{split}
&\gr{u}=\mathcal{A}\gr{U}+\mathcal{B}\gr{V}=(\A-3\B\D^{-1}\B)^{-1}(\gr{U}-3\B\D^{-1}\gr{V}),\\
&\gr{v}_1=\frac{2}{h}(\mathcal{B}^\top\gr{U}+3\mathcal{D}\gr{V})=\frac{6}{h}(\D-3\B\A^{-1}\B)^{-1}(\gr{V}-\B\A^{-1}\gr{U}),\\
&\gr{v}_2=\frac{h}{6}(3\mathcal{A}\gr{V}+\mathcal{B}\gr{W})=\frac{h}{2}(\A-3\B\D^{-1}\B)^{-1}(\gr{V}-\B\D^{-1}\gr{W}),\\
&\gr{w}=\mathcal{B}^\top\gr{V}+\mathcal{D}\gr{W}=(\D-3\B\A^{-1}\B)^{-1}(\gr{W}-3\B\A^{-1}\gr{V}),
\end{split}
\ee
the  compliance thermoelastic tensors. Contrarily to what could seem, the compliance thermoelastic tensors are only three because 
\be
\label{eq:legamev1v2}
\gr{v}_2=\frac{h^2}{12}\mathcal{A}\left[\mathcal{D}^{-1}\gr{v}_1+\frac{6}{h}\B(\A^{-1}\gr{U}-\D^{-1}\gr{W})\right].
\ee
In particular:
\begin{itemize}
\item $\gr{u}$ is the tensor of the coefficients of thermal expansion of the laminate in case of a uniform temperature change $t$; its SI units are $^\circ$C$^{-1}$;
\item $\gr{v}_1$ is the tensor of the coefficients of thermal expansion of the laminate due to a gradient of temperature $\nabla t$; its SI units are $(m\ ^\circ$C$)^{-1}$;
\item $\gr{v}_2$ is the tensor of the coefficients of thermal curvature of the laminate for a uniform change of temperature $t$; its SI units are $m\ ^\circ$C$^{-1}$;
\item $ \gr{w}$ is the tensor of the coefficients of thermal curvature of the laminate caused by a gradient of temperature $\nabla t$; its SI units are $^\circ$C$^{-1}$.
\end{itemize}

The symmetries of elasticity and of $\balpha$ ensure the symmetry of $\U,\V$ and $\W$, which in turn, through the minor symmetries of $\Ac,\Bc,\Dc$, gives the symmetries of $\u,\v_1,\v_2,\w$.

Concerning $\Bc$, we notice that in general  $\Bc\neq \Bc^\top$, i.e. it does not possess the major symmetries.  So, it has nine independent constants, as pointed out in  \cite{vannucci10ijss}. However, for some special  sequences, it can happen that $\Bc=\Bc^\top$, and even that it is rari-constant, cf. \cite{vannucci23a}.

We adopt the Kelvin's notation, \cite{kelvin,kelvin1}, for  2nd-rank tensors, e.g.:
\be
\N=\left\{
\begin{array}{c}
N_1=N_{11}\\
N_2=N_{22}\\
N_6=\sqrt{2}N_{12}
\end{array}
\right\},\ \
\ee
and similarly for $\M,\beps,\bk,\U,\V,\W,\u,\v_1,\v_2$ and $\w$, as well as for 4th-rank tensors, like
\be
\label{eq:kelvinmatrix}
\A\hspace{-1mm}=\hspace{-1mm}\left[
\begin{array}{ccc}
\A_{11}\hspace{-1mm}=\hspace{-1mm}\A_{1111}&\hspace{-1mm}\A_{12}\hspace{-1mm}=\hspace{-1mm}\A_{1122}&\hspace{-1mm}\A_{16}\hspace{-1mm}=\hspace{-1mm}\sqrt{2}\A_{1112}\\
\A_{12}\hspace{-1mm}=\hspace{-1mm}\A_{1122}&\hspace{-1mm}\A_{22}\hspace{-1mm}=\hspace{-1mm}\A_{2222}&\hspace{-1mm}\A_{26}\hspace{-1mm}=\hspace{-1mm}\sqrt{2}\A_{2212}\\
\A_{16}\hspace{-1mm}=\hspace{-1mm}\sqrt{2}\A_{1112}&\hspace{-1mm}\A_{26}\hspace{-1mm}=\hspace{-1mm}\sqrt{2}\A_{2212}&\hspace{-1mm}\A_{66}\hspace{-1mm}=\hspace{-1mm}2\A_{1212}
\end{array}
\right],
\ee
and the same for $\B,\D,\Ac,\Bc,\Dc$.

\section{Some basic elements of the polar method for laminates}
\begin{figure}[h]
\center
\includegraphics[width=.5\textwidth]{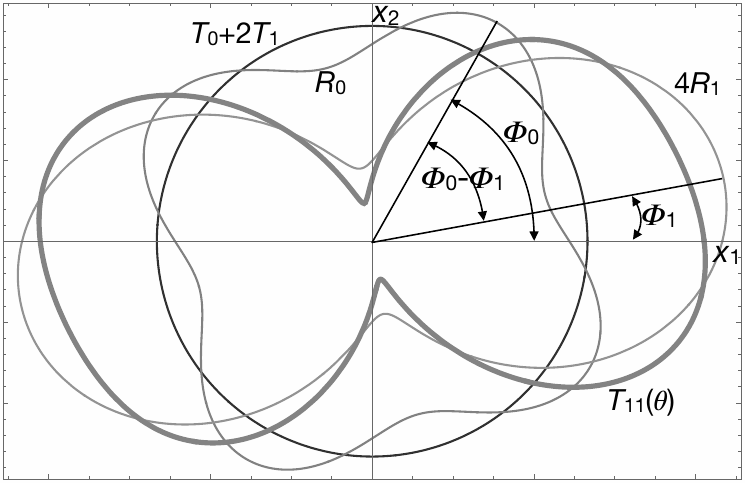}
\caption{The anisotropic phases of the component $\mathbb{T}_{11}(\theta)$ for a generic material.}
\label{fig:1}
\end{figure}

By the polar method, the Cartesian components in the Kelvin notation at a direction $\theta$ of a plane elastic tensor $\mathbb{T}$ are expressed as
\begin{equation}
\label{eq:mohr4}
{\begin{split}
&{\mathbb{T}_{11}{(}\theta{)}{=}{T}_{0}{+}{2}{T}_{1}{+}{R}_{0}\cos{4}\left({{{\varPhi}}_{0}{-}\theta}\right){+}{4}{R}_{1}\cos{2}\left({{{\varPhi}}_{1}{-}\theta}\right)},\\
&{\mathbb{T}_{12}{(}\theta{)}{=}{-}{T}_{0}{+}{2}{T}_{1}{-}{R}_{0}\cos{4}\left({{{\varPhi}}_{0}{-}\theta}\right)},\\
&\mathbb{T}_{16}(\theta)=\sqrt{2}\left[R_0\sin4\left(\varPhi_0-\theta\right)+2R_1\sin2\left(\varPhi_1-\theta\right)\right],\\
&{\mathbb{T}_{22}{(}\theta{)}{=}{T}_{0}{+}{2}{T}_{1}{+}{R}_{0}\cos{4}\left({{{\varPhi}}_{0}{-}\theta}\right){-}{4}{R}_{1}\cos{2}\left({{{\varPhi}}_{1}{-}\theta}\right)},\\
&{\mathbb{T}_{26}{(}\theta{)}{=}\sqrt{2}\left[{-}{R}_{0}\sin{4}\left({{{\varPhi}}_{0}{-}\theta}\right){+}{2}{R}_{1}\sin{2}\left({{{\varPhi}}_{1}{-}\theta}\right)\right]},\\
&{\mathbb{T}_{66}{(}\theta{)}{=}2\left[{T}_{0}{-}{R}_{0}\cos{4}\left({{{\varPhi}}_{0}{-}\theta}\right)\right]}.
\end{split}}
\end{equation}
$T_0,T_1,R_0,R_1$ are invariant moduli, and invariant is also the angle difference $\Phi_0-\Phi_1$.  Fixing one of the two polar angles amounts to fix a reference frame, normally $\Phi_1=0$. In the polar formalism all the elastic symmetries are determined by a peculiar value of an invariant: 
\begin{itemize}
\item ordinary orthotropy: $\Phi_0-\Phi_1=K\dfrac{\pi}{4}, \ K\in\{0,1\}$;
\item $R_0$-orthotropy: $R_0=0$, \cite{vannucci02joe};
\item square symmetry: ${R_1=0}$;
\item isotropy: $R_0=R_1=0$.
\end{itemize}
Hence, $T_0$ and $T_1$ are the {\it isotropy} and  $R_0,R_1$ and $\Phi_0-\Phi_1$  the {\it anisotropy invariants}. In practice, the polar method gives a decomposition of plane anisotropic elasticity, \cite{vannucci14mmas}, into an {\it isotropic phase}, depending on the two invariants $T_0$ and $T_1$, plus two {\it anisotropic phases}, shifted of the invariant angle $\Phi_0-\Phi_1$ and with amplitudes  proportional to $R_0$ and $R_1$ respectively; this decomposition is sketched in Fig. \ref{fig:1} for the component $\mathbb{T}_{11}$ of a generic material.

The above results are valid also for  $\A,\B,\D,\Ac,\Dc$ too, {\it but not for} $\Bc$, because it is not symmetric. 
We will denote by a superscript $A,B$ or $D$  a polar quantity of $\A,\B$ or $\D$ respectively; capital letters indicate stiffness tensors and lower-case letters   compliance tensors (e.g., $T_0^A,\Phi_1^A$ etc. indicate polar parameters of $\A$, while $t_0^A,\phi_1^A$ etc. the corresponding ones of $\Ac$, and similarly for $\B,\Bc,\D$ and $\Dc$). 

For what concerns a second-rank symmetric tensor {\bf L}, in the polar formalism it is
\be
\label{eq:mohr}
\begin{split}
&{L}_{1}(\theta)=T+R\cos2(\varPhi-\theta),\\
&{L}_{2}(\theta)=T-R\cos2(\varPhi-\theta),\\
&{L}_{6}(\theta)=\sqrt{2}R\sin2(\varPhi-\theta),
\end{split}
\ee
with $T, R$ two invariants, representing respectively the {\it isotropic} and the {\it anisotropic} phases of {\bf L}; $\Phi$ is an angle determined by the choice of the frame. In the following, we will indicate by 
$T^U,R^U,\Phi^U$ the polar components of $\U$, by $t^u,r^u,\phi^u$ those of $\u$ and similarly for $\V$, $\W$ and $\w$, while $t_1^v,r_1^v,\phi_1^v$ will denote the polar components of $\v_1$ and $t_2^v,r_2^v,\phi_2^v$ those of $\v_2$.

\section{Isotropic hybrid laminates composed of anisotropic plies}
Though never  used in practice, it is in principle possible to fabricate isotropic laminates stacking together anisotropic layers of different materials. This is actually the most general possible case and its interest, here, is in the fact that its analysis, exactly because concerning the most general case, allows to better understand the mathematical and mechanical aspects of the problem, also for the following, simpler cases.

For hybrid laminates, the polar components of $\A,\B$ and $\D$ are given by (a polar quantity of the basic layer has no superscript)
\begin{equation}
 \label{eq:compApolar}
 \A\ \ \rightarrow\ \
 \left\{
 \begin{split}
 &T_0^A=\frac{1}{h}\sum_{k=1}^n{T_0}_k(z_k-z_{k-1}),\\
 &T_1^A=\frac{1}{h}\sum_{k=1}^n{T_1}_k(z_k-z_{k-1}),\\
&R_0^Ae^{4i\Phi_0^A}=\frac{1}{h}\sum_{k=1}^n{R_0}_ke^{4i({\Phi_0}_k+\delta_k)}(z_k-z_{k-1}),\\
&R_1^Ae^{2i\Phi_1^A}=\frac{1}{h}\sum_{k=1}^n{R_1}_ke^{2i({\Phi_1}_k+\delta_k)}(z_k-z_{k-1}).
 \end{split}
 \right.
 \end{equation}
\begin{equation}
 \label{eq:compBpolar}
 \B\ \ \rightarrow\ \
 \left\{
 \begin{split}
 &T_0^B=\frac{1}{h^2}\sum_{k=1}^n{T_0}_k(z_k^2-z_{k-1}^2),\\
 &T_1^B=\frac{1}{h^2}\sum_{k=1}^n{T_1}_k(z_k^2-z_{k-1}^2),\\
&R_0^Be^{4i\Phi_0^B}=\frac{1}{h^2}\sum_{k=1}^n{R_0}_ke^{4i({\Phi_0}_k+\delta_k)}(z_k^2-z_{k-1}^2),\\
&R_1^Be^{2i\Phi_1^B}=\frac{1}{h^2}\sum_{k=1}^n{R_1}_ke^{2i({\Phi_1}_k+\delta_k)}(z_k^2-z_{k-1}^2).
 \end{split}
 \right.
 \end{equation}
\begin{equation}
 \label{eq:compDpolar}
 \D\ \ \rightarrow\ \
 \left\{
 \begin{split}
 &T_0^D=\frac{4}{h^3}\sum_{k=1}^n{T_0}_k(z_k^3-z_{k-1}^3),\\
 &T_1^D=\frac{4}{h^3}\sum_{k=1}^n{T_1}_k(z_k^3-z_{k-1}^3),\\
&R_0^De^{4i\Phi_0^D}=\frac{4}{h^3}\sum_{k=1}^n{R_0}_ke^{4i({\Phi_0}_k+\delta_k)}(z_k^3-z_{k-1}^3),\\
&R_1^De^{2i\Phi_1^D}=\frac{4}{h^3}\sum_{k=1}^n{R_1}_ke^{2i({\Phi_1}_k+\delta_k)}(z_k^3-z_{k-1}^3).
 \end{split}
 \right.
 \end{equation}
It is worth noting that the isotropic and anisotropic parts of all the tensors remain {\it separated} in the homogenization of the polar parameters, for all the tensors. Moreover, {\it special orthotropies are preserved}: 
 \begin{equation}
 \begin{split}
 &{R_0}_k=0\ \forall k\ \Rightarrow\ R_0^A=R_0^B=R_0^D=0,\\
 &{R_1}_k=0\ \forall k\ \Rightarrow\ R_1^A=R_1^B=R_1^D=0.
 \end{split}
 \end{equation}
 In particular, for  a laminate  composed by layers that are all square symmetric, though different,  $\A,\ \B$ and $\D$ are square symmetric too, no matter  the stacking sequence and  the layer orientations. The same is true for $R_0-$orthotropic plies, while ordinary orthotropy is not preserved through the homogenization process.

The polar parameters of $\U,\V,\W$ are
\be
\label{eq:compUpolar}
\gr{U}\ \rightarrow\ \left\{\begin{split}&T^U=\frac{1}{h}\sum_{k=1}^n{T_\gamma}_k(z_k-z_{k-1}),
\\&R^Ue^{2i\Phi^U}=\frac{1}{h}\sum_{k=1}^n{R_\gamma}_k e^{2i({\Phi_\gamma}_k+\delta_k)}(z_k-z_{k-1}),\end{split}\right.
\ee
\be
\label{eq:compVpolar}
\gr{V}\ \rightarrow\ \left\{\begin{split}&T^V=\frac{1}{h^2}\sum_{k=1}^n{T_\gamma}_k(z_k^2-z_{k-1}^2),
\\&R^Ve^{2i\Phi^V}=\frac{1}{h^2}\sum_{k=1}^n{R_\gamma}_ke^{2i({\Phi_\gamma}_k+\delta_k)}(z_k^2-z_{k-1}^2),\end{split}\right.
\ee
\be
\label{eq:compWpolar}
\gr{W}\ \rightarrow\ \left\{\begin{split}&T^W=\frac{4}{h^3}\sum_{k=1}^n{T_\gamma}_k(z_k^3-z_{k-1}^3),
\\&R^We^{2i\Phi^W}=\frac{4}{h^3}\sum_{k=1}^n{R_\gamma}_ke^{2i({\Phi_\gamma}_k+\delta_k)}(z_k^3-z_{k-1}^3),\end{split}\right.
\ee 
 where $T_\gamma,R_\gamma,\Phi_\gamma$ are the polar parameters of tensor $\bgamma(\delta_k)$.
 
 Let us suppose, now, that $\A$ and $\D$ are isotropic. This is possible if and only if
 \be
 \label{eq:condiso}
 R_0^A=R_1^A=R_0^D=R_1^D=0.
 \ee
 However, because plies are different, this implies nothing more; namely
 \begin{itemize}
 \item from eqs. (\ref{eq:compApolar})$_{1,2}$ and (\ref{eq:compDpolar})$_{1,2}$ we see that, in general,
 \be
 T_0^A
 \neq
 T_0^D,\ \ 
 T_1^A
 \neq
 T_1^D
 \Rightarrow\A\neq\D:
 \ee
though the extension and bending stiffnesses are isotropic, they are different;
 \item from eqs. (\ref{eq:compBpolar}) we get that, generally speaking,
 \be
 T_0^B\neq T_1^B,\ R_0^B\neq0,\  R_1^B\neq0: 
 \ee
$\B$ is not rari-constant ($\B_{12}\neq\B_{66}/2$), cf. Sect. \ref{sec:isolaminae}, nor  isotropic;
 \item because of these facts and eq. (\ref{eq:inversetensorsABD})$_{1,3}$, cf. \cite{vannucci23a}, in general it is
 \be
t_0^A\neq t_0^D,\ t_1^A\neq t_1^D,\ r_0^A\neq0,\ r_1^A\neq0,\ r_0^D\neq0,\ r_1^D\neq0\Rightarrow\Ac\neq\Dc
\ee
and also that $\Ac$ and $\Dc$ are not necessarily isotropic: as an effect of coupling the mechanical response in extension is not the same for intensity nor for material symmetries, in stiffness and in compliance, and the same is true for bending; this is a key point, that more that any other consideration shows the unconventional effects of coupling;
 \item from eq. (\ref{eq:inversetensorsABD})$_2$, we see that in general $\Bc\neq\Bc^\top$, so it depends upon nine independent parameters, cf. \cite{vannucci10ijss,vannucci23a}; in addition, $\Bc$ has not necessarily the same material symmetries of $\B$; hence, also if $\A$ and $\D$ are isotropic, coupling is, normally, anisotropic;
\item from eqs. (\ref{eq:compApolar})$_{4}$,  (\ref{eq:compDpolar})$_{4}$, (\ref{eq:compUpolar}) and (\ref{eq:compWpolar}) we get that, in general,
\be
R_1^A=0\nRightarrow R^U=0,\ R_1^D=0\nRightarrow R^W=0,\ T^U\neq T^W\Rightarrow\U\neq\W
\ee
 and also that $\U$ and $\W$ are not necessarily isotropic; in this general case, the isotropy of the elastic stiffness does not imply necessarily that of the thermoelastic one;
\item from eq. (\ref{eq:compVpolar}) we see that in general 
\be
T^V\neq0,\ R^V\neq0:
\ee
$\V$ is not isotropic nor purely anisotropic (i.e. with the isotropic part that vanishes);
\item from eqs. (\ref{eq:tenstherminverses})$_{1,4}$ and the previous considerations, we get that generally speaking
\be
 t^u\neq t^w,\ r^u\neq0,\ r^w\neq0:
 \ee
 $\u$ and $\w$ are not necessarily isotropic nor identical;
\item from eq. (\ref{eq:tenstherminverses})$_{2,3}$ and the results above, we have that in general
\be
t^v_1\neq0,\ t^v_2\neq0,\ r^v_1\neq0,\ r^v_2\neq0:
\ee
$\v_1$ and $\v_2$ are not isotropic nor purely anisotropic.
 \end{itemize}
 All these results have been  obtained putting eq. (\ref{eq:condiso}) into eqs. (\ref{eq:compApolar}) and (\ref{eq:compDpolar}) and then injecting eqs. (\ref{eq:compApolar}) to (\ref{eq:compWpolar}) into eqs. (\ref{eq:fundlawinversetherm}) and (\ref{eq:tenstherminverses}) after what the polar componetns of all the tensors can be computed through the converse of eqs. (\ref{eq:mohr4}) and (\ref{eq:mohr}), cf. \cite{Meccanica05,vannucci_libro}.  The results, obtained, thanks to a program for analytical computations, are omitted here because represented by rather long formulae.

 What is interesting to remark is exactly the fact that in general no real advantage arises from being $\A$ and $\D$ isotropic: coupling makes isotropy disappear from compliance tensors, so the deformation of the plate even under the action of isotropic extension forces or bending moments is not isotropic: in this completely general set of laminates, isotropy does not give a real advantage and it is questionable to affirm that the isotropy of $\A$ and $\D$ really gives the isotropy of the behavior of the laminate. More interesting situations are obtained for the next sets of laminates, where some conditions are relaxed and simpler situations obtained.

\section{Isotropic hybrid laminates composed of isotropic plies}
A simple way to obtain isotropy is to use isotropic layers. This is typically the case of plates obtained bonding together metallic layers, but not only. The most important application of this type of laminates is bimetal plates, that are analyzed in the next Section. Here, we want to put in evidence the general characteristics of a laminate composed of different isotropic layers.

The isotropy of each layer enforces that of all the tensors; this can be easily seen by the polar formalism, because, eqs. (\ref{eq:compApolar})$_{3,4}$  to (\ref{eq:compDpolar})$_{3,4}$,
\be
{R_0}_k=0\ \forall k\Rightarrow R_0^A=R_0^B=R_0^D=0,\ {R_1}_k=0\ \forall k\Rightarrow R_1^A=R_1^B=R_1^D=0,
\ee
i.e. $\A,\B$ and $\D$ are automatically isotropic. However, it is still $\A\neq\D$, because, in general, eqs. (\ref{eq:compApolar})$_{1,2}$ and (\ref{eq:compDpolar})$_{1,2}$, $T_0^A\neq T_0^D,T_1^A\neq T_1^D$. 

Similar considerations can be made for the thermoelastic tensors $\U,\V,\W$. In fact, $\Q$ and $\balpha$ are isotropic for each ply, which implies that $\bgamma$ is isotropic too. As a consequence, by eqs. (\ref{eq:compApolar}) to (\ref{eq:compDpolar})
\be
{R_\gamma}_k=0\ \forall k\Rightarrow R^U=R^V=R^W=0,
\ee
i.e. $\U,\V$ and $\W$ are isotropic as well.

Following the computational steps described in the previous Section, we can  calculate all the compliance tensors; the whole set of results is detailed hereafter in terms of polar parameters:
\begin{itemize}
\item tensor $\Ac$: 
\be
\label{eq:polarAhybrid}
t_0^A= \frac{T_0^D}{4\left(T_0^AT_0^D-3{T_0^B}^2\right)},\ t_1^A=\frac{T_1^D}{16\left(T_1^AT_1^D-3{T_1^B}^2\right)};
\ee
\item tensor $\Bc$:
\be
\label{eq:polarBhybrid}
t_0^B= -\frac{3T_0^B}{4\left(T_0^AT_0^D-3{T_0^B}^2\right)},\ t_1^B=-\frac{3T_1^B}{16\left(T_1^AT_1^D-3{T_1^B}^2\right)};
\ee
\item tensor $\Dc$:
\be
\label{eq:polarDhybrid}
t_0^D= \frac{T_0^A}{4\left(T_0^AT_0^D-3{T_0^B}^2\right)},\ t_1^D=\frac{T_1^A}{16\left(T_1^AT_1^D-3{T_1^B}^2\right)};
\ee
\item tensors $\u,\w$:
\be
\label{eq:polaruwhybrid}
t^u=\frac{3T_1^B T^V-T_1^D T^U}{4\left(T_1^A T_1^D-3 {T_1^B}^2\right)},\
t^w=-\frac{3T_1^B T^V-T_1^A T^W}{4\left(T_1^A T_1^D-3 {T_1^B}^2\right)};
\ee
\item tensors $\v_1,\v_2$:
\be
\label{eq:polarv1v2hybrid}
t^v_1=\frac{3}{2h}\frac{T_1^A T^V-T_1^B T^U}{T_1^A T_1^D-3 {T_1^B}^2},\
t^v_2=\frac{h}{8}\frac{T_1^D T^V-T_1^B T^W}{T_1^A T_1^D-3 {T_1^B}^2}.
\ee
\end{itemize}

To reconstruct the Cartesian matrices representing the different tensors is very easy, once the above quantities known: by eq. (\ref{eq:mohr4}) we see that each 4th-rank tensor ($\theta$ can be ignored, all the tensors are  isotropic) is of the form
\be
\label{eq:isogeneral}
\mathbb{T}=\left[\begin{array}{ccc}T_0+2T_1&-T_0+2T_1&0\\-T_0+2T_1&T_0+2T_1&0\\0&0&2T_0\end{array}\right],
\ee
while all the 2nd-rank tensors are of the type, cf. eq. (\ref{eq:mohr}),
\be
\label{eq:iso2general}
\mathbf{L}=\left\{\begin{array}{c}T\\T\\0\end{array}\right\}.
\ee

The  ratios
\be
\frac{t^w}{t^u}=-\frac{T_1^A T^W-3 T_1^B T^V}{T_1^D T^U-3 T_1^B T^V},\ \frac{t^v_2}{t^v_1}=\frac{h^2}{12}\frac{T_1^B T^W-T_1^D T^V}{T_1^B T^U-T_1^A T^V},
\ee
 measure, respectively, the importance of the bending thermal response with respect to the extension one and the relative importance of the two thermoelastic coupling tensors $\v_1$ and $\v_2$.
 
Incidentally, because $T_0^A,T_1^A,T_0^D,T_1^D$, as well as $t_0^A,t_1^A,t_0^D,t_1^D$ are positive quantities, cf. \cite{vannucci15ijss,vannucci23}, from the above relations we get the following bounds for the thermoelastic invariants:
\be
T_0^AT_0^D-3{T_0^B}^2>0,\ T_1^AT_1^D-3{T_1^B}^2>0,\ T_1^DT^U<3T_1^BT^V<T_1^AT^W.
\ee

It is worth noting that $\B$ is not rari-constant ($\B_{12}\neq\B_{66}/2$), see Sect. \ref{sec:isolaminae}, and that 
\be
\det\B=16{T_0^B}^2T_1^B\neq0.
\ee
Moreover, $\Bc=\Bc^\top$ and 
\be
\det\Bc=-\frac{27{T_0^B}^2T_1^B}{16\left(T_0^AT_0^D-3{T_0^B}^2\right)^2\left(T_1^AT_1^D-3{T_1^B}^2\right)}\neq0.
\ee
Hence, the isotropy of all the layers ensures the symmetry of $\Bc$ and the non singularity of $\B$ and $\Bc$.

Finally, we see that in this case the isotropy of each layer is a sufficient condition to ensure the isotropy of all the elastic and thermoelastic tensors. 

Let us consider now a laminate, free to deform under the action of some internal forces or changes of temperature. We want to see what is the overall deformation of the plate. First, let us consider the case of a laminate submitted to only in-plane actions $\N$; then, from eqs. (\ref{eq:fundlawinversetherm}) and (\ref{eq:isogeneral}) we have that
\be
\beps=\frac{1}{h}\Ac\N=\frac{1}{h}\left\{\begin{array}{c}t_0^A(N_1-N_2)+2t_1^A(N_1+N_2)\\t_0^A(N_2-N_1)+2t_1^A(N_1+N_2)\\2t_0^AN_6\end{array}\right\},
\ee
\be
\label{eq:curvatura1}
\bk=\frac{2}{h^2}\Bc^\top\N=\frac{2}{h^2}\left\{\begin{array}{c}t_0^B(N_1-N_2)+2t_1^B(N_1+N_2)\\t_0^B(N_2-N_1)+2t_1^B(N_1+N_2)\\2t_0^BN_6\end{array}\right\}.
\ee
Then, through eq.  (\ref{eq:polarAhybrid}) and (\ref{eq:polarBhybrid}), we can express the deformation as a function of the stiffness polar components:
\be
\beps=\frac{1}{4h}\left\{\begin{array}{c}\frac{T_0^D(N_1-N_2)}{T_0^AT_0^D-3{T_0^B}^2}+\frac{T_1^D(N_1+N_2)}{2\left(T_1^AT_1^D-3{T_1^B}^2\right)}\\\frac{T_0^D(N_2-N_1)}{T_0^AT_0^D-3{T_0^B}^2}+\frac{T_1^D(N_1+N_2)}{2\left(T_1^AT_1^D-3{T_1^B}^2\right)}\\\frac{2T_0^DN_6}{T_0^AT_0^D-3{T_0^B}^2}\end{array}\right\},
\ee
\be
\bk=-\frac{3}{2h^2}\left\{\begin{array}{c}\frac{T_0^B(N_1-N_2)}{T_0^AT_0^D-3{T_0^B}^2}+\frac{T_1^B(N_1+N_2)}{2\left(T_1^AT_1^D-3{T_1^B}^2\right)}\\\frac{T_0^B(N_2-N_1)}{T_0^AT_0^D-3{T_0^B}^2}+\frac{T_1^B(N_1+N_2)}{2\left(T_1^AT_1^D-3{T_1^B}^2\right)}\\\frac{2T_0^BN_6}{T_0^AT_0^D-3{T_0^B}^2}\end{array}\right\}.
\ee
If $N_1=N_2,N_6=0$, e.g. a square plate stretched or compressed equally along the edges, then $\beps$ and $\bk$ are purely spherical: the plate deforms homothetically in its plane and curves like the surface of a sphere.  

If  $N_2=-N_1,N_6=0$, e.g. a square plate equally stretched  along two opposite sides and compressed along the two others, then
\be
\beps=\frac{1}{2h}\frac{T_0^D}{T_0^AT_0^D-3{T_0^B}^2}\left\{\begin{array}{c}1\\-1\\0\end{array}\right\}N_1,\ 
\bk=-\frac{3}{h^2}\frac{T_0^B}{T_0^AT_0^D-3{T_0^B}^2}\left\{\begin{array}{c}1\\-1\\0\end{array}\right\}N_1.
\ee
So, in this case  $\beps$ and $\bk$ are purely deviatoric: the plate deforms in its plane without changing its surface and bends with opposite curvatures, so that the deformed surface is made of hyperbolic points, because the Gauss curvature $K=\kappa_1\kappa_2$ is negative, and it is a minimal surface, because the mean curvature $H=(\kappa_1+\kappa_2)/2$ is null, cf. \cite{vannucci_alg,pressley}.

If the plate is acted upon only by couples, i.e. $\N=\gr{O},\M\neq\gr{O}$, then proceeding in the same way we get
\be
\beps=\frac{2}{h^2}\Bc\M=-\frac{3}{2h^2}\left\{\begin{array}{c}\frac{T_0^B(\M_1-\M_2)}{T_0^AT_0^D-3{T_0^B}^2}+\frac{T_1^B(\M_1+\M_2)}{2\left(T_1^AT_1^D-3{T_1^B}^2\right)}\\\frac{T_0^B(\M_2-\M_1)}{T_0^AT_0^D-3{T_0^B}^2}+\frac{T_1^B(\M_1+\M_2)}{2\left(T_1^AT_1^D-3{T_1^B}^2\right)}\\\frac{2T_0^B\M_6}{T_0^AT_0^D-3{T_0^B}^2}\end{array}\right\},
\ee
\be
\bk=\frac{12}{h^3}\Dc\M=\frac{3}{h^3}\left\{\begin{array}{c}\frac{T_0^A(\M_1-\M_2)}{T_0^AT_0^D-3{T_0^B}^2}+\frac{T_1^A(\M_1+\M_2)}{2\left(T_1^AT_1^D-3{T_1^B}^2\right)}\\\frac{T_0^A(\M_2-\M_1)}{T_0^AT_0^D-3{T_0^B}^2}+\frac{T_1^A(\M_1+\M_2)}{2\left(T_1^AT_1^D-3{T_1^B}^2\right)}\\\frac{2T_0^A\M_6}{T_0^AT_0^D-3{T_0^B}^2}\end{array}\right\}.
\ee
Once more, if $\M_1=\M_2,\M_6=0,\ \beps$ and $\bk$ are purely spherical, while if $\M_2=-\M_1,\M_6=0$, they are purely deviatoric, with the same observations for what concerns the deformation of the laminate.

If now only  a change of temperature $t$ acts on the same plate,  from eqs. (\ref{eq:fundlawinversetherm}), (\ref{eq:polaruwhybrid}), (\ref{eq:polarv1v2hybrid}) and (\ref{eq:iso2general}) we get, 
\be
\beps=t\ \u=t\left\{\begin{array}{c}t^u\\t^u\\0\end{array}\right\}=\frac{3\ T_1^B T^V-T_1^D T^U}{4\left(T_1^A T_1^D-3 {T_1^B}^2\right)}\left\{\begin{array}{c}1\\1\\0\end{array}\right\}t,
\ee
\be
\label{eq:curvatura2}
\bk=t\ \v_1=t\left\{\begin{array}{c}t^v_1\\t^v_1\\0\end{array}\right\}=\frac{3}{2h}\frac{T_1^A T^V-T_1^B T^U}{T_1^A T_1^D-3 {T_1^B}^2}\left\{\begin{array}{c}1\\1\\0\end{array}\right\}t;
\ee
also in this case,  $\beps$ and $\bk$ are purely spherical.

\section{Bimetal plates}
\begin{figure}[h]
\center
\includegraphics[width=.5\textwidth]{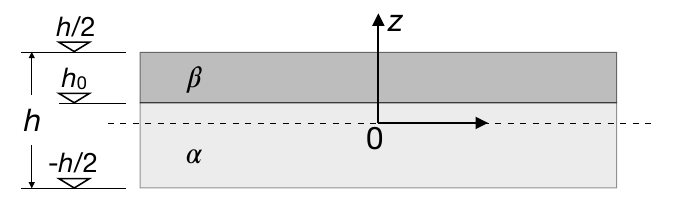}
\caption{General scheme of a bimetal plate.}
\label{fig:2}
\end{figure}

The most important application of the previous case is that of bimetal plates. In such a case two layers only, of different metallic alloys and not necessarily of the same thickness, are superposed to form a plate. Because the stack is particularly simple, it is possible to explicit the formulae of the previous Section as functions of the material properties of the two layers. Let us refer to the scheme of Fig. \ref{fig:2}: the plate is composed by two isotropic layers, denoted by $\alpha$ and $\beta$, $h_0$ is the position of the interface. Applying to this case eqs. (\ref{eq:compApolar}) to (\ref{eq:compWpolar}) it is simple to get
\be
\label{eq:bimetalinit}
T_0^A=\frac{T_0^\alpha+T_0^\beta}{2}+(T_0^\alpha-T_0^\beta)\frac{h_0}{h},\ T_1^A=\frac{T_1^\alpha+T_1^\beta}{2}+(T_1^\alpha-T_1^\beta)\frac{h_0}{h},\medskip
\ee
\be
T_0^B=(T_0^\alpha-T_0^\beta)\left(\frac{h_0^2}{h^2}-\frac{1}{4}\right),\ T_1^A=(T_1^\alpha-T_1^\beta)\left(\frac{h_0^2}{h^2}-\frac{1}{4}\right),\medskip
\ee
\be
T_0^D=\frac{T_0^\alpha+T_0^\beta}{2}+4(T_0^\alpha-T_0^\beta)\frac{h_0^3}{h^3},\ T_1^D=\frac{T_1^\alpha+T_1^\beta}{2}+4(T_1^\alpha-T_1^\beta)\frac{h_0^3}{h^3},\medskip
\ee
\be
T^U\hspace{-1mm}=\frac{T_\gamma^\alpha+T_\gamma^\beta}{2}+(T_\gamma^\alpha-T_\gamma^\beta)\frac{h_0}{h},\ 
T^V\hspace{-1mm}=(T_\gamma^\alpha-T_\gamma^\beta)\left(\frac{h_0^2}{h^2}-\frac{1}{4}\right),\ 
T^W\hspace{-1mm}=\frac{T_\gamma^\alpha+T_\gamma^\beta}{2}+4(T_\gamma^\alpha-T_\gamma^\beta)\frac{h_0^3}{h^3}.
\ee
Replacing these values into eqs. (\ref{eq:polarAhybrid}) to (\ref{eq:polarv1v2hybrid}) gives the polar components of all the compliance tensors:

\be
\begin{array}{c}
t_0^A=\dfrac{4}{\psi_0} \left[8 h {h_0}^3 ({T_0^\alpha }-{T_0^\beta })+h^4 ({T_0^\alpha }+{T_0^\beta })\right],\bigskip\\
t_1^A=\dfrac{1}{\psi_1}\left[8 h {h_0}^3 ({T_1^\alpha }-{T_1^\beta })+h^4 ({T_1^\alpha }+{T_1^\beta })\right],
\end{array}
\ee
\be
\begin{array}{c}
t_0^B=\dfrac{6}{\psi_0} h^2 \left(h^2-4 {h_0}^2\right) ({T_0^\alpha }-{T_0^\beta }),\bigskip\\
t_1^B=\dfrac{3}{2\psi_1} h^2 \left(h^2-4 {h_0}^2\right) ({T_1^\alpha }-{T_1^\beta }),
\end{array}
\ee
\be
\begin{array}{c}
t_0^D=
\dfrac{4}{\psi_0} h^3 \left[2 {h_0} ({T_0^\alpha }-{T_0^\beta })+h ({T_0^\alpha }+{T_0^\beta })\right],\bigskip\\
t_1^D=\dfrac{1}{\psi_1}h^3 \left[2 {h_0} ({T_1^\alpha }-{T_1^\beta })+h ({T_1^\alpha }+{T_1^\beta }))\right],
\end{array}
\ee
with
\be
\begin{split}
&\psi_0=2 (h+2 {h_0})^4
{T_0^\alpha }^2+4 (h^2-4 {h_0}^2) \left(7 h^2+4 {h_0}^2\right) {T_0^\alpha } {T_0^\beta }+2 (h-2 {h_0})^4
{T_0^\beta }^2,\\
&\psi_1=2 (h+2 {h_0})^4
{T_1^\alpha }^2+4 (h^2-4 {h_0}^2) \left(7 h^2+4 {h_0}^2\right) {T_1^\alpha } {T_1^\beta }+2 (h-2 {h_0})^4
{T_1^\beta }^2.
\end{split}
\ee
For the thermoelastic tensors, we get
\be
\begin{array}{c}
t^u=\dfrac{1}{2\psi_1}\left\{
(h+2 h_0) \left[(h+2 h_0)^3 {T_1^\alpha }+(h-2 h_0) \left(7 h^2+8 h h_0+4 h_0^2\right) {T_1^\beta
}\right] {T_\gamma ^\alpha }+\right.\\
\left.+(h-2 h_0) \left[(h-2 h_0)^3 {T_1^\beta
}+(h+2 h_0) \left(7 h^2-8 h h_0+4 h_0^2\right) {T_1^\alpha }\right] {T_\gamma ^\beta }\right\}\bigskip\\
t^w=\dfrac{1}{2\psi_1}\left\{
(h+2 h_0) \left[(h+2 h_0)^3 {T_1^\alpha }+(h-2 h_0) \left(7 h^2-8 h h_0+4 h_0^2\right) {T_1^\beta
}\right] {T_\gamma ^\alpha }+\right.\\
\left.+(h-2 h_0) \left[(h-2 h_0)^3 {T_1^\beta
}+(h+2 h_0) \left(7 h^2+8 h h_0+4 h_0^2\right) {T_1^\alpha }\right] {T_\gamma ^\beta }\right\}
\end{array}
\ee
and
\be
\label{eq:bimetalfin}
t^v_1=\dfrac{12}{\psi_1} h (h^2-4 h_0^2) (T_1^\alpha T_\gamma^\beta-T_1^\beta T_\gamma^\alpha),\ 
t^v_2=\dfrac{1}{\psi_1} h^3 (h^2-4 h_0^2) (T_1^\alpha T_\gamma^\beta-T_1^\beta T_\gamma^\alpha).
\ee
It is worth noting that 
\be
\frac{t_2^v}{t_1^v}=\frac{h^2}{12},
\ee
a result valid also for laminates composed by identical anisotropic layers, see the next Section.

Particularly interesting is the case of layers of equal thickness, i.e. of $h_0=0$:
\be
t_0^A=t_0^D= \frac{2 (T_0^\alpha+T_0^\beta)}{{T_0^\alpha}^2+14 T_0^\alpha T_0^\beta+{T_0^\beta}^2},\ 
t_1^A=t_1^D= \frac{T_1^\alpha+T_1^\beta}{2\left({T_1^\alpha}^2+14 T_1^\alpha T_1^\beta+{T_1^\beta}^2\right)},
\ee
\be
t_0^B= \frac{3 (T_0^\alpha-T_0^\beta)}{{T_0^\alpha}^2+14 T_0^\alpha T_0^\beta+{T_0^\beta}^2},\ 
t_1^B=  \frac{3 (T_1^\alpha-T_1^\beta)}{4\left({T_1^\alpha}^2+14 T_1^\alpha T_1^\beta+{T_1^\beta}^2\right)},
\ee
\be
t^u=t^w=\frac{ T_1^\beta (7 T_\gamma^\alpha+T_\gamma^\beta)+T_1^\alpha (T_\gamma^\alpha+7 T_\gamma^\beta)}{4\left({T_1^\alpha}^2+14 T_1^\alpha T_1^\beta+{T_1^\beta}^2\right)},
\ee
\be
t_1^v=\frac{6}{h}\frac{ T_1^\alpha T_\gamma^\beta-T_1^\beta T_\gamma^\alpha}{ {T_1^\alpha}^2+14 T_1^\alpha T_1^\beta+{T_1^\beta}^2},\ 
 t_2^v=\frac{h}{2}\frac{ T_1^\alpha T_\gamma^\beta-T_1^\beta T_\gamma^\alpha}{ {T_1^\alpha}^2+14 T_1^\alpha T_1^\beta+{T_1^\beta}^2}.
\ee
It is worth noting that in this particular case, $\A=\D,\Ac=\Dc,\U=\V,\u=\w$: the bimetal with equal thicknesses is a {\it quasi-homogeneous coupled laminate}  (QHCL), i.e. a laminate with identical behavior in extension and in bending, cf. \cite{vannucci01ijss,CompScTech01,vannucci_libro}. Actually, this is the only  known general case of hybrid QHCL (this result is not valid if $h_0\neq0$).

 \section{Isotropic laminates made of identical anisotropic plies}
 \label{sec:isolaminae}

We consider now the case of laminates composed by identical layers; in particular, we assume that the basic layer is orthotropic with  a material frame fixed by the choice $\Phi_1=0$, so that, eq. (\ref{eq:mohr4})
\begin{equation}
\label{eq:Q}
\begin{split}
&\mathbb{Q}_{11}{(}\theta{)}{=}{T}_{0}{+}{2}{T}_{1}{+}(-1)^K{R}_{0}\cos{4}\theta{+}{4}{R}_{1}\cos{2}\theta,\\
&\mathbb{Q}_{12}{(}\theta{)}{=}{-}{T}_{0}{+}{2}{T}_{1}{-}(-1)^K{R}_{0}\cos{4}\theta,\\
&\mathbb{Q}_{16}(\theta)=-\sqrt{2}\left[(-1)^KR_0\sin4\theta+2R_1\sin2\theta\right],\\
&\mathbb{Q}_{22}{(}\theta{)}{=}{T}_{0}{+}{2}{T}_{1}{+}(-1)^K{R}_{0}\cos{4}\theta{-}{4}{R}_{1}\cos{2}\theta,\\
&\mathbb{Q}_{26}{(}\theta{)}{=}\sqrt{2}\left[(-1)^K{R}_{0}\sin{4}\theta{-}{2}{R}_{1}\sin{2}\theta\right],\\
&\mathbb{Q}_{66}{(}\theta{)}{=}2\left[{T}_{0}{-}(-1)^K{R}_{0}\cos{4}\theta\right].
\end{split}
\end{equation}
Phisycally, this implies that  the tensor of thermal expansion coefficients $\balpha$ is  orthotropic, or isotropic, with the same orthotropy axes of $\Q$, so that
\be
\Phi_\gamma=\Phi_1+\lambda\frac{\pi}{2},\ \ \lambda\in\{0,1\}.
\ee

For  laminates composed of identical plies, a very important case for applications, the  relations (\ref{eq:compApolar}) to (\ref{eq:compDpolar}) for $\A,\B,\D,$ become
 \begin{equation}
  \label{eq:compABDpolareq}
  \begin{split}
&\mathbb{A}\rightarrow\left\{
\begin{array}{l}
T_0^A=T_0,\\
T_1^A=T_1,\\
R_0^Ae^{4i\varPhi_0^A}={R_0e^{4i\varPhi_0}}{\left(\xi_1+i\xi_2\right)},\\
R_1^Ae^{2i\varPhi_1^A}={R_1e^{2i\varPhi_1}}{\left(\xi_3+i\xi_4\right)};
\end{array}\right.\\
&\mathbb{B}\rightarrow\left\{
\begin{array}{l}
T_0^B=0,\\
T_1^B=0,\\
R_0^Be^{4i\varPhi_0^B}={R_0e^{4i\varPhi_0}}{\left(\xi_5+i\xi_6\right)},\\
R_1^Be^{2i\varPhi_1^B}={R_1e^{2i\varPhi_1}}{\left(\xi_7+i\xi_8\right)};
\end{array}\right.\\
&\mathbb{D}\rightarrow\left\{
\begin{array}{l}
T_0^D=T_0,\\
T_1^D=T_1,\\
R_0^De^{4i\varPhi_0^D}={R_0e^{4i\varPhi_0}}(\xi_{9}+i\xi_{10}),\\
R_1^De^{2i\varPhi_1^D}={R_1e^{2i\varPhi_1}}(\xi_{11}+i\xi_{12}),
\end{array}\right.
\end{split}
\end{equation}
and those for $\U,\V$ and $\W$, eqs. (\ref{eq:compUpolar}) to (\ref{eq:compWpolar}),
\be
\label{eq:UVWpolaridlayers}
\begin{split}
&\gr{U}\ \rightarrow\ \left\{\begin{split}&T^U=T_\gamma,\\&R^Ue^{2i\Phi^U}=R_\gamma e^{2i\Phi_\gamma}(\xi_3+i\xi_4),\end{split}\right.\\
&\gr{V}\ \rightarrow\ \left\{\begin{split}&T^V=0,\\&R^Ve^{2i\Phi^V}=R_\gamma e^{2i\Phi_\gamma}(\xi_7+i\xi_8);\end{split}\right.\\
&\gr{W}\ \rightarrow\ \left\{\begin{split}&T^W=T_\gamma,\\&R^We^{2i\Phi^W}=R_\gamma e^{2i\Phi_\gamma}(\xi_{11}+i\xi_{12}).\end{split}\right.
\end{split}
\ee
In the above equations, the quantities $\xi_i,\ i=1,...,12,$ are the so-called {\it lamination parameters} \cite{TsaiHahn,vannucci_libro},   quantities accounting for the geometry of the stacking sequence (i.e. for orientations and position of the layers on the stack):
\begin{equation}
\label{eq:laminationparameters}
\begin{aligned}
&{{{{\xi}}_{1}{{+}}{i}{{\xi}}_{2}}{{=}}\mathop{\sum}\limits_{{k}{{=}}{1}}\limits^{n}a_k\hspace{0.33em}{{e}^{4i{{\delta}}_{k}}}},\hspace{1cm}
& &{{{{\xi}}_{3}{{+}}{i}{{\xi}}_{4}}{{=}}\mathop{\sum}\limits_{{k}{{=}}{1}}\limits^{n}a_k\hspace{0.33em}{{e}^{2i{{\delta}}_{k}}}},\\
&{{{{\xi}}_{5}{{+}}{i}{{\xi}}_{6}}{{=}}\mathop{\sum}\limits_{{k}{{=}}{1}}\limits^{n}{{b}_{k}\hspace{0.33em}{e}^{4i{{\delta}}_{k}}}},\hspace{1cm}
& &{{{{\xi}}_{7}{{+}}{i}{{\xi}}_{8}}{{=}}\mathop{\sum}\limits_{{k}{{=}}{1}}\limits^{n}{{b}_{k}\hspace{0.33em}{e}^{2i{{\delta}}_{k}}}},\\
&{{{{\xi}}_{9}{{+}}{i}{{\xi}}_{10}}{{=}}\mathop{\sum}\limits_{{k}{{=}}{1}}\limits^{n}{{d}_{k}\hspace{0.33em}{e}^{4i{{\delta}}_{k}}}},\hspace{1cm}
& &{{{{\xi}}_{11}{{+}}{i}{{\xi}}_{12}}{{=}}\mathop{\sum}\limits_{{k}{{=}}{1}}\limits^{n}{{d}_{k}\hspace{0.33em}{e}^{2i{{\delta}}_{k}}}},\\%
\end{aligned}
\end{equation}
with
\begin{equation}
\label{eq:coefABCD}
a_k=\frac{1}{n},\ b_k=\frac{1}{n^2}(2k-n-1),\ d_k=\frac{1}{n^3}\left[12k(k-n-1)+4+3n(n+2)\right].
\end{equation}

So, through the polar formalism we can see that {\it for laminates made of identical layers}:
\begin{itemize}
\item the isotropic part of $\A$ and $\D$ is equal to that of the basic layer: $T_0^A=T_0^D=T_0,T_1^A=T_1^D=T_1$;
\item in the same way, the isotropic part of $\U$ and $\W$ is equal to that of $\bgamma:T^U=T^W=T_\gamma$;
\item $\B$ is exclusively anisotropic, i.e. its isotropic part is null, $T_0^B=T_1^B=0$, so its average on $\{0,2\pi\}$  is zero;
\item $\V$ is exclusively anisotropic, i.e. its isotropic part is null, $T^V=0$, so also its average on  $\{0,2\pi\}$  is zero.
\end{itemize}

Tensors $\B$ and $\Bc$ are indeed special operators. We have already seen that in general $\Bc\neq\Bc^\top$. Concerning $\B$, if the plies are identical,  $T_0^B=T_1^B$ (and in addition they are null), which gives that $\B$ is a {\it rari-constant tensor}, \cite{vannucci2016}: it has also the Cauchy-Poisson symmetries, \cite{pearson,love,benvenuto91}, besides the minor and major ones. In practice,  $\B$ is completely symmetric with respect to any permutation of the indices, i.e. it is also
\be
\B_{1122}=\B_{1212}\ \rightarrow\ \B_{12}=\frac{\B_{66}}{2}.
\ee
In some special cases, it possible that not only $\Bc=\Bc^\top$, but also that $\Bc_{1122}=\Bc_{1212}$, i.e. that also $\Bc$ is rari-constant, cf. \cite{vannucci23a}.

So, for this special set of laminates, the stiffness behavior in extension is equal to that in bending, $\A=\D$: these laminates belong to the class  QHCL; in addition, they just correspond to the mean (isotropic phase) of the reduced stiffness tensor $\Q$ of the basic layer. The same is true for the thermoelastic stiffness tensors: $\U=\V$ and they are the mean part of $\bgamma$.

Several results have been proved in \cite{vannucci23a,vannucci23b} for QHCL:
\begin{itemize}
\item $\A=\B,\Rightarrow\Ac=\Dc$: quasi-homogeneity is valid also for the compliance tensors,
\item $\U=\V\Rightarrow\u=\w$: the same is true for the thermoelastic behavior;
\item $\Bc=\Bc^\top$: the compliance coupling tensor is symmetric too;
\end{itemize}

Because $T_0^B=T_1^B=T^V=0$, and putting $\theta=\Phi_1^B$ to fix a frame,  from eqs. (\ref{eq:mohr4})   and (\ref{eq:mohr}) we get
\begin{equation}
\label{eq:Bsimplified}
\begin{split}
&\B_{11}=R_0^B\cos4\varPhi_B+4R_1^B,\\
&\B_{12}=-R_0^B\cos4\varPhi_B,\\
&\B_{16}=\sqrt{2}R_0^B\sin4\varPhi_B,\\
&\B_{22}=R_0^B\cos4\varPhi_B-4R_1^B,\\
&\B_{26}=-\sqrt{2}R_0^B\sin4\varPhi_B,\\
&\B_{66}=-2R_0^B\cos4\varPhi_B,
\end{split}
\end{equation}
with  $\Phi_B=\Phi_0^B-\Phi_1^B$, an invariant of $\B$.

Concerning the thermoelastic coupling tensor $\V$, it has been shown in \cite{vannucci23b} that it can be put in the form

\be
\label{eq:formuleV}
\V=(-1)^\lambda\rho\ R_1^B\left\{\begin{array}{c}1\\-1\\0\end{array}\right\},
\ee
with
\be
\label{eq:rho}
\rho=\frac{R^V}{R_1^B}=\frac{R_\gamma}{R_1},
\ee
a material parameter that characterizes at the same time the layer and the laminate.

The calculation of the compliance parameters in the most general case of $\B$ completely anisotropic is possible, but the results are very long. That is why, for the sake of simplicity, we give here the results for the case of $\B$ orthotropic, i.e. for  
\be\Phi_B=K\frac{\pi}{4},\ K\in\{0,1\}.\ee
 With these parameters, we get the following results for the compliance tensors:
\be
\label{eq:resultsinit}
\begin{array}{c}
t_0^A=t_0^D=\dfrac{\left(3 {R_1^B}^2-T_0 T_1\right) \left[6 {R_1^B}^2 T_0-\left(T_0^2-3 {R_0^B}^2\right) T_1\right]}{4 \eta \left(3 {R_0^B}^2-T_0^2\right)},\bigskip\\
t_1^A=t_1^D= \dfrac{6 {R_1^B}^2 T_0-\left(T_0^2-3 {R_0^B}^2\right) T_1}{16\eta},\bigskip\\
r_0^A=r_0^D=\dfrac{3}{4}{R_1^B}^2\left|\dfrac{6 {R_1^B}^2 T_0-\left(T_0^2+3 {R_0^B}^2\right) T_1}{ \eta\left(3 {R_0^B}^2-T_0^2\right)}\right|,\bigskip\\
r_1^A=r_1^D= \dfrac{3}{8}\dfrac{R_0^BR_1^B T_1}{ |\eta|},\ 
\phi_0^A=\phi_0^D=\phi_1^A=\phi_1^D=0,\bigskip
\end{array}
\ee
\be
\begin{array}{c}
t_0^B=(-1)^K\dfrac{9 R_0^B {R_1^B}^2 \left(3 {R_1^B}^2-T_0 T_1\right)}{2\eta \left(3 {R_0^B}^2-T_0^2\right)},\ 
t_1^B=(-1)^K\dfrac{9 R_0^B {R_1^B}^2}{8 \eta},\bigskip\\
r_0^B=\dfrac{3}{4}R_0^B\left|\dfrac{T_1\left[6 {R_1^B}^2 T_0-\left(T_0^2-3 {R_0^B}^2\right) T_1\right]-18{R_1^B}^4}{\eta \left(3 {R_0^B}^2-T_0^2\right)}\right|,\bigskip\\
r_1^B=\dfrac{3}{8}\left|\dfrac{R_1^B T_0 T_1-6 {R_1^B}^3}{ \eta}\right|,\ 
\phi_0^B=\phi_1^B=0,\bigskip
\end{array}
\ee
\be
\label{eq:thermoconstantes}
\begin{array}{c}
t^u=t^w=\dfrac{6 {R_1^B}^2 T_0 (T_\gamma+(-1)^\lambda T_1\ \rho)-\left(T_0^2-3 {R_0^B}^2\right) T_1 T_\gamma-36 (-1)^\lambda {R_1^B}^4 \rho}{4\eta},\bigskip\\
r^u=r^w=\dfrac{3}{2}{R_0^B} {R_1^B} T_1 \left|\dfrac{T_\gamma-(-1)^\lambda T_1\ \rho}{\eta}\right|,\ \phi^u=\phi^w=0,\bigskip
\end{array}
\ee
\be
\begin{array}{c}
t^v_1=\dfrac{9}{h}\dfrac{R_0^B{R_1^B}^2  (T_\gamma-(-1)^\lambda T_1\ \rho)}{\eta},\bigskip\\
r^v_1=\dfrac{3}{h} {R_1^B}  \left|\dfrac{\left(T_0T_1-6{R_1^B}^2\right)(T_\gamma-(-1)^\lambda T_1\ \rho)}{\eta}\right|,\ \phi^v_1=0,\bigskip
\end{array}
\ee
\be
\label{eq:resultsfin}
\begin{array}{c}
t^v_2=\dfrac{3}{4}h\dfrac{R_0^B{R_1^B}^2  (T_\gamma-(-1)^\lambda T_1\ \rho)}{\eta},\bigskip\\
r^v_2=\dfrac{h}{4} {R_1^B}  \left|\dfrac{\left(T_0T_1-6{R_1^B}^2\right)(T_\gamma-(-1)^\lambda T_1\ \rho)}{\eta}\right|,\ \phi^v_2=0,\bigskip
\end{array}
\ee
with
\be
\eta=12 {R_1^B}^2 T_0 T_1+\left(3 {R_0^B}^2-T_0^2\right) T_1^2-36 {R_1^B}^4.
\ee

Some  remarks can be done:
\begin{itemize}
\item as a consequence of coupling, though $\A=\D$ are isotropic, $\Ac=\Dc$ are not isotropic; in the case of $\Phi_B=0$ they are $k=0$-orthotropic, otherwise they can be completely anisotropic as well;
\item the same is true also for the thermoelastic tensors: $\U=\W$ are isotropic but $\u=\w$ are orthotropic;
\item unlike $\B$ and $\V$, the isotropic part of $\Bc$ and $\v_1,\v_2$ is not null, despite the fact that the layers are identical; in small words, $\B$ makes appear the anisotropic part for $\Ac=\Dc,\u=\w$ and the isotropic one for $\Bc,\v_1,\v_2$;
\item changing the type of ordinary orthotropy for $\B$, i.e. changing $K=0$ with $K=1$, has the only effect of changing the sign of $t_0^B$ and $t _1^B$, the isotropic part of $\Bc$;  
\item the result that $\Bc=\Bc^\top$ for $\A=\D$ isotropic is due to the orthotropy of $\B$; in case of $\B$ completely anisotropic, $\Bc\neq\Bc^\top$; 
\item we get the ratios 
\be
\frac{t^v_2}{t^v_1}=\frac{r^v_2}{r^v_1}=\frac{h^2}{12},
\ee
which means that (the result is valid also for completely anisotropic $\B$, i.e. if $\Phi_B\neq0$)
\be
\v_2=\frac{h^2}{12}\v_1;
\ee
\item remembering eq. (\ref{eq:rho}), the condition $T_\gamma-(-1)^\lambda T_1\ \rho=0$ can be rewritten as
\be
\label{eq:condthemisotcomp}
R_1T_\gamma-(-1)^\lambda T_1R_\gamma=0;
\ee
this equation is automatically satisfied by the use of plies reinforced by balanced fabrics, i.e. with $R_1=0$; in fact, we have seen that $R_1=0\Rightarrow R_\gamma=0$. This is a sufficient condition to have  $r^u=r^w=0$, i.e. isotropic $\u=\w$, and $t^v_1=t^v_2=r^v_1=r^v_2=0$, i.e. the vanishing of $\v_1$ and $\v_2$. There is another possibility for satisfying eq. (\ref{eq:condthemisotcomp}): it is a particular combination of  the parameters $T_1,R_1,T_\gamma$ and $R_\gamma$. Though possible in principle, this combination, that actually depends exclusively on the material and not on the stacking sequence, just like the other one $R_1=0$, is more theoretical than practical. However, physically it is possible to get coupled laminates with $\V=\v_1=\v_2=\gr{O}$ and isotropic $\U=\W$ and $\u=\w$ just by the use of a particular material, independently from the stacking sequence;
\item it is
\be
\det\B=32R_0^B{R_1^B}^2,\ \det\Bc=-\frac{27R_0^B{R_1^B}^2}{8\eta\left(3{R_0^B}^2-T_0^2\right)}:
\ee
$\B$ and $\Bc$ are hence non singular.
\end{itemize}

An interesting question is: how to obtain isotropic QHCL laminates? The most simple way is to join two strategies: the first one, is the rule of Werren and Norris, \cite{werren53}, which ensures the isotropy of $\A$; this rule consists in putting an equal number of layers at  least three directions, each one shifted of an equal angle. Then, if this rule is applied to the set of  {\it quasi-trivial} QHCL, \cite{vannucci01ijss,CompScTech01,vannucci_libro}, quasi-homogeneity automatically ensures $\D=\A$, so if $\A$ is isotropic, also $\D$ will be isotropic. Quasi-trivial solutions are particular sequences, that ensure a given property (quasi-homogeneity, uncoupling) if all the layers belonging to certain sub-groups of layer are all put at the same orientation, no matter  the value of this one. This strategy has been used in \cite{CompStruct02} to obtain {\it fully isotropic} laminates, i.e. laminates with $\A=\D$ isotropic and $\B=\mathbb{O}$. 
 If isotropic QHCL are sought for, the strategy must be applied to the set of quasi-trivial QHCL; some solutions of this type to which the Werren and Norris rule has been applied is given in Tab. \ref{tab:1}.
 \begin{table}[h]
\caption{Some Werren and Norris solutions for isotropic QHCL plates.}
\begin{small}
\begin{center}
\begin{tabular}{cc}
\toprule
&$\left[0^\circ\  45^\circ\  -45^\circ\  0^\circ\  -45^\circ\  90^\circ\   0^\circ\  45^\circ\  45^\circ\  90^\circ\   0^\circ\  -45^\circ\  90^\circ\   -45^\circ\  45^\circ\  90^\circ  \right],$ \\
&$\left[0^\circ\ 45^\circ\  45^\circ\  0^\circ\ 45^\circ\  0^\circ\  0^\circ\  45^\circ\  -45^\circ\ 90^\circ\ 90^\circ\ -45^\circ\ 90^\circ\  -45^\circ\ -45^\circ\ 90^\circ \right],$ \\
16&$\left[0^\circ\ 45^\circ\  45^\circ\  90^\circ\  45^\circ\  0^\circ\ 0^\circ\ 45^\circ\  -45^\circ\ 90^\circ\  90^\circ\  -45^\circ\ 0^\circ\ -45^\circ\ -45^\circ\ 90^\circ \right],$ \\
plies&$\left[0^\circ\ 45^\circ\  45^\circ\  90^\circ\  45^\circ\  0^\circ\ 90^\circ\  45^\circ\  -45^\circ\ 0^\circ\ 90^\circ\  -45^\circ\ 0^\circ\ -45^\circ\ -45^\circ\ 90^\circ \right],$ \\
&$\left[0^\circ\ -45^\circ\ 45^\circ\  0^\circ\ -45^\circ\ 0^\circ\ 0^\circ\ -45^\circ\ 45^\circ\ 90^\circ\ 90^\circ\ 45^\circ\  90^\circ\ -45^\circ\ 45^\circ\ 90^\circ \right],$\\
&$\left[0^\circ\  -45^\circ\  45^\circ\  90^\circ\  -45^\circ\  0^\circ\  0^\circ\  -45^\circ\  45^\circ\  90^\circ\  90^\circ\  45^\circ\  0^\circ\  -45^\circ\  45^\circ\  90^\circ   \right],$ \\
\midrule
&$\left[0^\circ\ 0^\circ\ 60^\circ\   0^\circ\ 60^\circ\   60^\circ\   -60^\circ\   0^\circ\ 0^\circ\ 0^\circ\ 60^\circ\   -60^\circ\   60^\circ\   -60^\circ\   -60^\circ\   -60^\circ\   -60^\circ\   60^\circ  \right],$\\
&$\left[0^\circ\ 60^\circ\   0^\circ\ -60^\circ\   0^\circ\ 0^\circ\ 60^\circ\   -60^\circ\   -60^\circ\   60^\circ\   0^\circ\ 60^\circ\   0^\circ\ 60^\circ\   -60^\circ\   -60^\circ\   -60^\circ\   60^\circ  \right],$\\
18&$\left[0^\circ\ 60^\circ\   -60^\circ\   0^\circ\ 60^\circ\   60^\circ\   -60^\circ\   0^\circ\ 60^\circ\   60^\circ\   0^\circ\ -60^\circ\   0^\circ\ -60^\circ\   -60^\circ\   0^\circ\ -60^\circ\   60^\circ  \right],$\\
plies&$\left[0^\circ\ 60^\circ\   60^\circ\   -60^\circ\   60^\circ\   0^\circ\ 60^\circ\   0^\circ\ -60^\circ\   0^\circ\ 0^\circ\ 60^\circ\   -60^\circ\   -60^\circ\   60^\circ\   -60^\circ\   -60^\circ\   0^\circ  \right],$\\
&$\left[0^\circ\ 60^\circ\   -60^\circ\   -60^\circ\   60^\circ\   0^\circ\ 60^\circ\   0^\circ\ -60^\circ\   60^\circ\   0^\circ\ 60^\circ\   -60^\circ\   -60^\circ\   0^\circ\ 0^\circ\ -60^\circ\   60^\circ  \right],$\\
&$\left[0^\circ\ 60^\circ\   -60^\circ\   -60^\circ\   60^\circ\   60^\circ\   -60^\circ\   0^\circ\ 60^\circ\   60^\circ\   0^\circ\ -60^\circ\   0^\circ\ -60^\circ\   0^\circ\ 0^\circ\ -60^\circ\   60^\circ  \right],$\\
 \bottomrule
 \end{tabular}
\end{center}
 \end{small}
\label{tab:1}
\end{table}

 To remark that a fully isotropic laminate, i.e. with isotropic $\A=\D$ and $\B=\mathbb{O}$, has a minimum of 18 plies, cf. \cite{CompStruct02}, while in Tab. \ref{tab:1}  six isotropic QHCL sequences with 16 plies are presented (there are 32 solutions of this type with 16 plies) and six with 18 plies (on the whole, they are 296). Isotropic QHCL laminates with  less than 16 plies do not exist in the set of quasi-trivial solutions, but they can be found using numerical optimization techniques, like those described in \cite{vannucci06,vannucci07,vannucci09algo,vannucci13jota,MAMS13,vincenti10,ahmadian11}.
 
 We consider in the following some special cases of coupled isotropic laminates composed of identical plies.
 
 \subsection{Coupled laminates with $\V=\gr{O}$}
From eqs. (\ref{eq:compABDpolareq}), (\ref{eq:UVWpolaridlayers}) and (\ref{eq:laminationparameters}) we can see that 
\be
R_1^B=0\Rightarrow\xi_8+i\xi_8=0\Rightarrow R^V=0:
\ee
if $\B$ is square symmetric, i.e. if $R_1^B=0$, then $\V=\gr{O}$: the thermoelastic coupling vanishes, but not the elastic one: $\B\neq\mathbb{O}$, because $R_0^B\neq0$. Hence, there can exist laminates that are elastically coupled, but  without thermal coupling, at least for what concerns stiffness. The existence of this type of laminates has been investigated in \cite{vannucci12joe1,vannucci23b}. 

The simplest way to obtain such laminates, is to use a square symmetric basic layer, i.e. having $R_1=R_\gamma=0$, which is the case of layers reinforced by balanced fabrics, i.e. having the same amount of fibers in warp and weft. In fact, eqs. (\ref{eq:compABDpolareq}) and (\ref{eq:UVWpolaridlayers}),
\be
R_1=0\Rightarrow R_1^A=R_1^B=R_1^D=0,\ R_\gamma=0\Rightarrow R^U=R^V=R^W=0.
\ee
To remark that the material square symmetry of a layer is a sufficient condition to get at the same time $R_1=0$, i.e. $\Q$ square symmetric, and $\balpha$ isotropic, which gives necessarily $\bgamma$ isotropic, i.e. $R_\gamma=0$.

Another advantage of using square symmetric layers is that it is easier to get the isotropy of $\A$: it is sufficient to put the same number of plies in two directions shifted of $\pi/4$; this is sufficient to get also $R_0^A=0$, and hence the isotropy of $\A$. If the already described strategy of using quasi-trivial stacks of QHCL is used, we get also $\D=\A$. Because only two orientations are needed in this case, the number of possible solutions is much higher than with unidirectional plies. In particular, all the antisymmetric stacks, i.e. sequences of the type $[0^\circ_p,45^\circ_p]$, with $p=n/2$, are laminates of this type and, in addition, they have the highest degree of coupling, as defined in \cite{vannucci01joe}.

Putting $R_1^B=R_\gamma=0$, the results (\ref{eq:resultsinit}) to (\ref{eq:resultsfin}) become:
\be
\label{eq:complianceV0}
\begin{array}{c}
t_0^A=t_0^D=\dfrac{T_0}{4\left(T_0^2-3{R_0^B}^2\right)},\  t_1^A=t_1^D=\dfrac{1}{16T_1},\ r_0^A=r_0^D=0,\ r_1^A=r_1^D=0,\bigskip\\
t_0^B=t_1^B=0,\ r_0^B=\dfrac{3R_0^B}{4\left(T_0^2-3{R_0^B}^2\right)},\ r_1^B=0,\bigskip\\
t^u=t^w=\dfrac{T_\gamma}{4T_1},\ r^u=r^w=0,\ t^v_1=t^v_2=0,\ r^v_1=r^v_2=0.
\end{array}
\ee
Hence, in this case also $\Ac=\Dc$ are isotropic, as well as $\u=\w$. More interesting, $\v_1=\v_2=\gr{O}$: thermoelastic coupling is null also in compliance. This is a case of {\it extension- and warp-free thermally stable laminates}, \cite{vannucci12joe1}, i.e., of coupled laminates that are completely insensitive to the thermal coupling. This category of laminates is particularly interesting for applications: laminates fabricated using pre-pregs plies are cured at high temperature and then cooled down to the room temperature, i.e. they are submitted to a severe change of temperature $t<0$. The fact that they are warp-free, i.e. that $\v_1=\gr{O}$, ensures that the plate is not bent by $t$, so it conserves its shape, despite the fact that it is coupled. 

Because $t_0^A$ and $t_0^D$ are positive quantities, we get also the bound
\be
T_0^4-3{R_0^B}^2>0.
\ee

We can ponder whether or not other types of coupled isotropic laminates with isotropic $\Ac=\Dc$ can exist; this question has been  considered, at least for the elastic part, in \cite{vannucci23b}, in the most general case; the answer is no: the only condition for having also the isotropy of the compliances is to have $R_1^B=0$. This can be seen also by eqs. (\ref{eq:resultsinit})$_{3,4}$: the only way to have $r_0^A=r_0^D=0$ and $r_1^A=r_1^D=0$ is to have $R_1^B=0$. 

For the thermoelastic part, this is sufficient too, but two other conditions, apparently, exist, cf. (\ref{eq:thermoconstantes})$_2$:
\be
\label{eq:condizioniisotropiatermcompl}
R_0^B=0,\ T_\gamma-(-1)^\lambda T_1\ \rho=0.
\ee
The first condition is examined in the next Section; the second one has  been considered above and  actually is redundant. In fact, we have already remarked above that  the use of plies reinforced by balanced fabrics, i.e. with $R_1=0$, automatically ensures not only that $R_1^B=0$, but also the second condition in eq. (\ref{eq:condizioniisotropiatermcompl}). If $R_1\neq0$ but $R_1^B=0$,  eq. (\ref{eq:condizioniisotropiatermcompl})$_2$ may be unsatisfied, but anyway it will be $r^u=r^w=0$ just because $R_1^B=0$. Of course, the other possibility ensuring the isotropy of $\u=\w$, but {\it not} of $\Ac=\Dc$, is a combination of the parameters $T_1,R_1,T_\gamma$ and $R_\gamma$ that satisfies the last equation above, but, as already said in the previous Section, this combination is more theoretical than practical. 

 Finally, the only condition to obtain the isotropy of all the elastic and thermoelastic compliance tensors is $R_1^B=0$.
Besides the use of plies reinforced by balanced fabrics, i.e. with $R_1=0$, we can consider the following question: what are the stacking sequences of unidirectional layers ($R_0\neq0, R_1\neq0$) that ensure $R_0^A=R_1^A=R_1^B=R_0^D=R_1^D=0$? In particular,  can these solutions be found in the set of quasi-trivial QHCL laminates, like those in Tab. \ref{tab:1}, once applied the rule of Werren and Norris for the isotropy of $\A$ (and hence of $\D$, by quasi-homogeneity)? To this purpose, among these solutions one has to look for those having {\it also} $R_1^B=0$, so, if $R_1\neq0$, satisfying, cf. eq. (\ref{eq:laminationparameters})$_4$:
\be
{{{{\xi}}_{7}{{+}}{i}{{\xi}}_{8}}{{=}}\mathop{\sum}\limits_{{k}{{=}}{1}}\limits^{n}{{b}_{k}\hspace{0.33em}{e}^{2i{{\delta}}_{k}}}}=0.
\ee
Through eq. (\ref{eq:coefABCD})$_2$, we get the condition
\be
\begin{split}
&\sum_{k=1}^n(2k-n-1)(\cos2\delta_k+i\sin2\delta_k)=\\
&=2\sum_{k=1}^nk(\cos2\delta_k+i\sin2\delta_k)-(n+1)\sum_{k=1}^n\cos2\delta_k+i\sin2\delta_k=0.
\end{split}
\ee
The last summation above is null whenever a Werren and Norris sequence is used, because 
\be
R_1^A=0\iff\xi_3+i\xi_4=\frac{1}{n}\sum_{k=1}^n\cos2\delta_k+i\sin2\delta_k=0.
\ee
Hence, to have $R_1^B=0$ with a quasi trivial QHCL sequence of the Werren and Norris type the condition is
\be
\sum_{k=1}^nk(\cos2\delta_k+i\sin2\delta_k)=0.
\ee
It is apparent that such a condition can never be satisfied (unless for  stacks  uncoupled independently from the orientations, i.e. of the  quasi-trivial type, e.g. symmetric; of course, this is not relevant for coupled plates, studied here). In fact, the coefficient $k$ increases from 1 to $n$ and if a solution existed, this no longer should be one when the numbering of the layers were inverted, which is also possible, or, in small words, if the plate were put upside down.  Of course, this is paradoxical, which means that solutions of this type cannot exist. 

However, it has been proven in \cite{vannucci01ijss} that the quasi-trivial set is not the whole set of quasi homogeneous laminates; other solutions   having isotropic $\A=\D$ and $\B\neq\mathbb{O}$ can hopefully  exist, to be found by numerical approaches, but until now no one of them is known. Finally, the best way to have coupled laminates with $\V=\gr{O}$ is to use layers reinforced by balanced fabrics, which is indeed a very common practice.

\subsection{Coupled laminates with $R_0^B=0$}
The case of $R_0^B$ is rather interesting, for different reasons. For the elastic part, the problem has already been studied in \cite{vannucci23a}, we complete here the results with also the thermoelastic part:
\be
\label{eq:casoRB0}
\begin{array}{c}
t_0^A=t_0^D=\dfrac{T_0T_1-3{R_1^B}^2}{4T_0\left(T_0T_1-6{R_1^B}^2\right)},\
 t_1^A=t_1^D=\dfrac{T_0}{16\left(T_0T_1-6{R_1^B}^2\right)},\bigskip\\
r_0^A=r_0^D=\dfrac{3{R_1^B}^2}{4T_0\left(T_0T_1-6{R_1^B}^2\right)},\
r_1^A=r_1^D=0,\ \phi_0^A=\phi_0^D=0,\bigskip\\
t_0^B=t_1^B=r_0^B=0,\ 
r_1^B=\dfrac{3{R_1^B}}{8\left(T_0T_1-6{R_1^B}^2\right)},\ \phi_1^B=0,\bigskip\\
t^u=t^w= \dfrac{T_0 T_\gamma-6 (-1)^\lambda {R_1^B}^2 \rho}{4 \left(T_0 T_1-6 {R_1^B}^2\right)},\ r^u=r^w=0,\bigskip\\
t^v_1=t^v_2=0,\ r^v_1=\dfrac{3}{h}R_1^B\dfrac{\left|T_\gamma-(-1)^\lambda T_1\ \rho\right|}{T_0 T_1-6 {R_1^B}^2},\
r^v_2=\dfrac{h}{4}R_1^B\dfrac{\left|T_\gamma-(-1)^\lambda T_1\ \rho\right|}{T_0 T_1-6 {R_1^B}^2}.
\end{array}
\ee
We then remark that:
\begin{itemize}
\item $\Ac$ and $\Dc$ are square symmetric;
\item $\Bc$ is $R_0$-orthotropic, like $\B$;
\item $\u$ and $\w$ are isotropic;
\item $\v_1$ and $\v_2$ are purely anisotropic;
\item because $t_0^A, t_0^D,t_1^A,t_1^D,t^u$ and $t^w$ are positive quantities, we get the following bounds for the invariants:
\be
T_0T_1-6{R_1^B}^2>0,\ T_0 T_\gamma-6 (-1)^\lambda {R_1^B}^2 \rho>0.
\ee
\end{itemize}
To obtain $R_0^B=0$ for coupled isotropic laminates is much easier than obtaining $R_1^B=0$. A first strategy is to use $R_0$-orthotropic layers. These can be obtained reinforcing an isotropic matrix with a balanced fabric but having the warp and weft fibers rotated of $\pi/4$ instead that orthogonal, \cite{vannucci02joe}. Because such materials are not frequent, $R_0^B=0$ can be easily get also using unidirectional layers: from eq. (\ref{eq:laminationparameters}) we see that 
\be
R_0^B=0\iff\xi_5+i\xi_6=\sum_{k=1}^nb_k(\cos4\delta_k+i\sin4\delta_k)=0.
\ee
It is rather easy to find a quasi-trivial QHCL solution of the  Werren and Norris type with four orientations, $\{0^\circ, 45^\circ,-45^\circ, 90^\circ\}$, that  satisfies this equation, e.g. all the 16-ply sequences in Tab. \ref{tab:1} are of this type.

\section{Numerical examples}
\subsection{Example 1: bimetal plate}
\begin{figure}[h]
\center
\includegraphics[width=.3\textwidth]{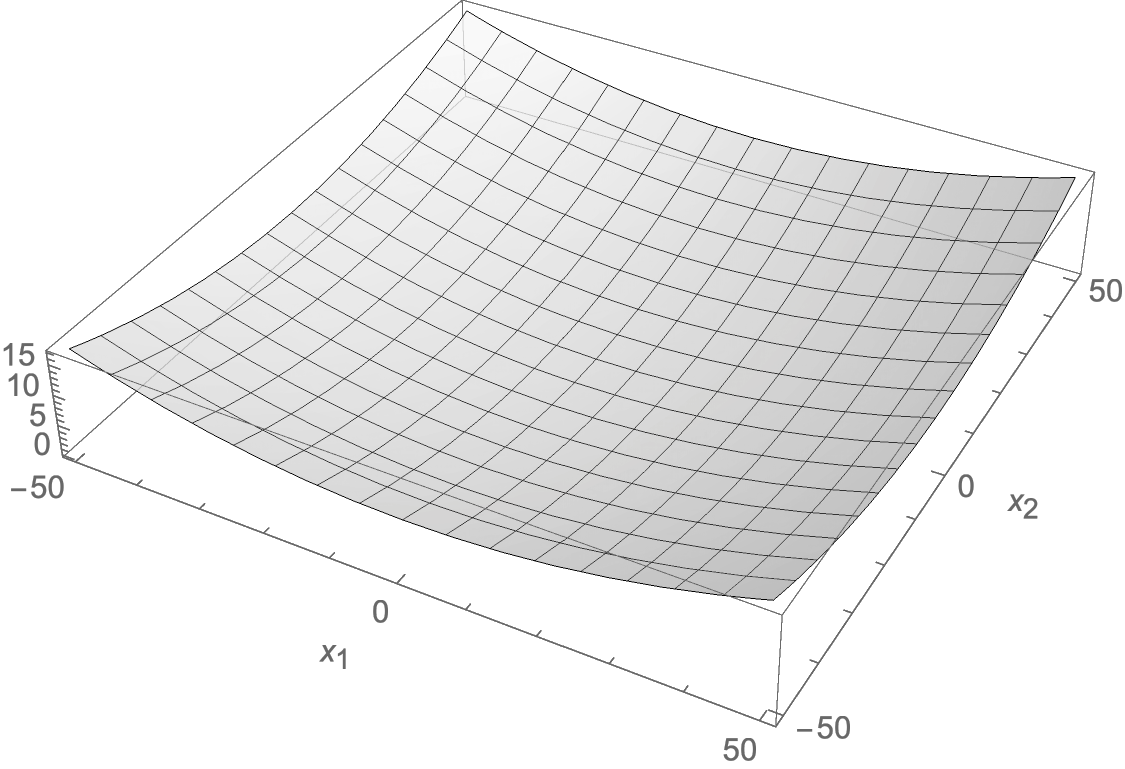}
\includegraphics[width=.3\textwidth]{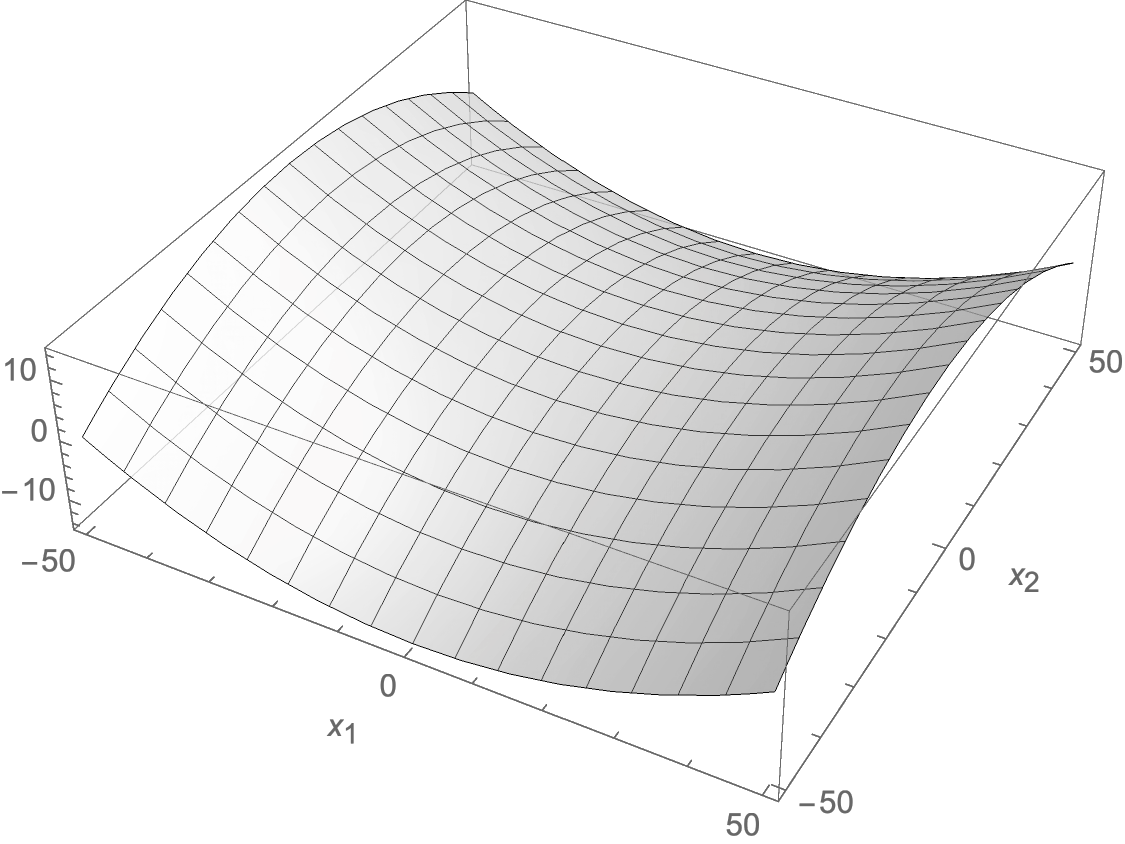}
\includegraphics[width=.3\textwidth]{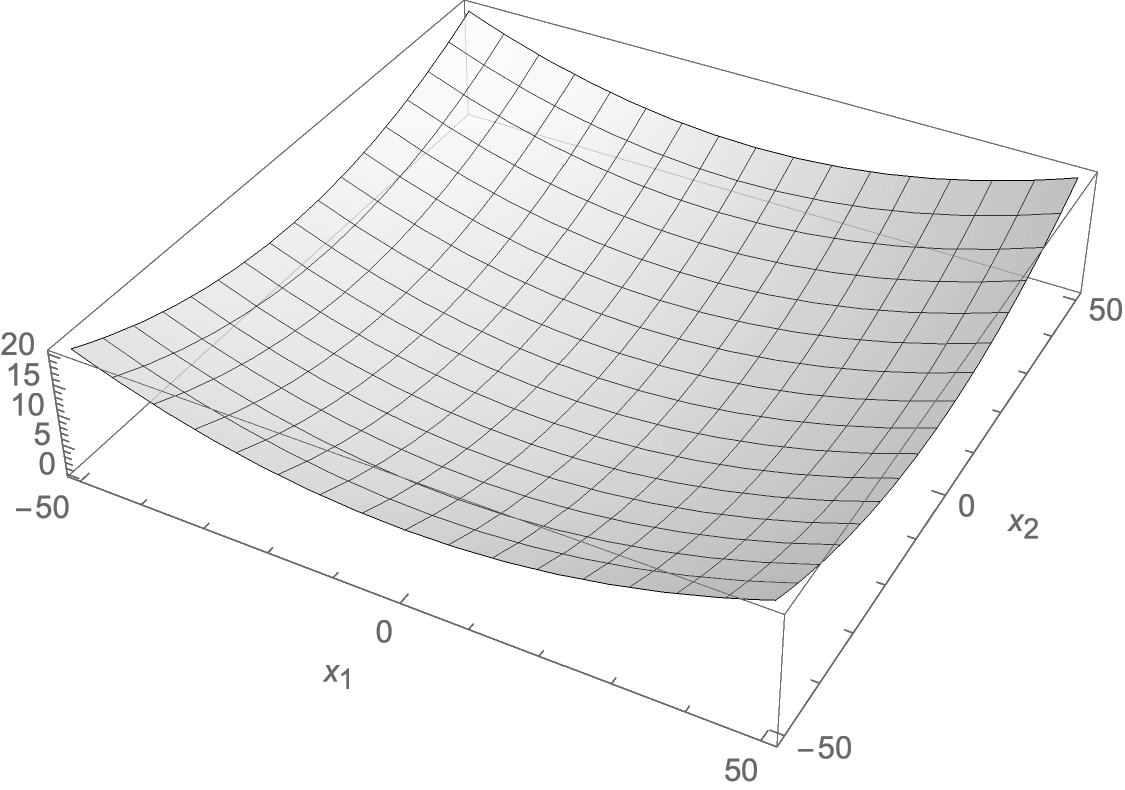}
\caption{Deformation due to coupling of the bimetal example. From the left the cases: $N_1=N_2,N_2=-N_1$, with $N_1=1000$ N/mm, and $t=50^\circ$C. The effects are magnified 10 times.}
\label{fig:3}
\end{figure}

Let us consider a bimetal plate, composed of layer of steel and one of aluminium, with thickness of 2 mm and 1 mm respectively. For an isotropic material, \cite{vannucci_libro},
\be
\Q_{11}=\Q_{22}=\frac{E}{1-\nu^2},\ \Q_{12}=\frac{\nu\ E}{1-\nu^2},\ \Q_{66}=\frac{E}{1+\nu},\ \balpha=\alpha\ \gr{I},
\ee
with $E$ the Young's modulus, $\nu$ the Poisson's coefficient and $\alpha$ the thermal expansion coefficient. This gives:
\be
T_0=\frac{E}{4(1-\nu^2)}(2-\nu-\nu^2),\ T_1=\frac{E}{4(1-\nu)},\ T_\gamma=(\Q_{11}+\Q_{12})\alpha=\frac{\alpha\ E}{1-\nu}.
\ee
With this formulae, we get the  values in Tab. \ref{tab:2} for the two materials.
\begin{table}[h]
\caption{Characteristics of the materials  of the bimetal plate.}
\begin{small}
\begin{center}
\begin{tabular}{lcccccc}
\toprule
Material&$E$&$\nu$&$\alpha$&$T_0$&$T_1$&$T_\gamma$\\
 &\small{MPa}&&\small{$^\circ$C$^{-1}$}&\small{MPa}&\small{MPa}&\small{MPa $^\circ$C$^{-1}$}\\
\midrule
Steel&210000&0.30&$1.2\times10^{-5}$&92885&75000&3.6\\
Aluminium&70000&0.33&$2.4\times10^{-5}$&30658&26119&2.5\\
 \bottomrule
 \end{tabular}
\end{center}
 \end{small}
\label{tab:2}
\end{table}
The parameters of the bimetal can be calculated through eqs. (\ref{eq:bimetalinit}) to (\ref{eq:bimetalfin}), they are presented in Tab. \ref{tab:3}.
\begin{table}[h]
\caption{Parameters of the bimetal plate.}
\begin{small}
\begin{center}
\begin{tabular}{ccccccccccc}
\toprule
$T_0^A$&$T_1^A$&$T_0^B$&$T_1^B$&$T_0^D$&$T_1^D$&&&$T^U$&$T^V$&$T^W$\\
\cmidrule{1-6}\cmidrule{9-11}
\multicolumn{6}{c}{\scriptsize{MPa$\times10^4$}}&&&\multicolumn{3}{c}{\scriptsize{MPa $^\circ$C$^{-1}$}}\\
\midrule
7.21&5.87&-1.38&-1.09&6.29&5.15&&&3.23&-0.24&3.07\\
\midrule
$t_0^A$&$t_1^A$&$t_0^B$&$t_1^B$&$t_0^D$&$t_1^D$&&$t^u$&$t_1^v$&$t_2^v$&$t^w$\\
\cmidrule{1-6}\cmidrule{8-11}
\multicolumn{6}{c}{\scriptsize{MPa$^{-1}\times10^{-6}$}}&&\scriptsize{$^\circ$C$^{-1}$}&\scriptsize{(mm $^\circ$C)$^{-1}$}&\scriptsize{mm $^\circ$C$^{-1}$}&\scriptsize{$^\circ$C$^{-1}$}\\
\midrule
3.97&1.20&2.61&0.76&4.55&1.37&&1.49$\times10^{-5}$&3.92$\times10^{-6}$&2.94$\times10^{-6}$&1.62$\times10^{-5}$\\
 \bottomrule
 \end{tabular}
\end{center}
 \end{small}
\label{tab:3}
\end{table}
Through eqs. (\ref{eq:curvatura1}) and (\ref{eq:curvatura2}) we can evaluate the effects of coupling due, respectively, to in-plane forces and a temperature change $t$. We present such effects in Fig. \ref{fig:3}, for the cases of membrane forces $N_1=N_2, N_2=-N_1$, with $N_1=1000$ N/mm, and for a temperature change $t=50^\circ$C on a square plate 100 mm wide.

\subsection{Example 2: laminate with $R_1^B=0$}
The simplest case is that of a laminate composed of just two square symmetric layers turned of $\pi/4$.  Such a plate is thermally stable, i.e. $\V=\v_1=\v_2=\gr{O}$: no thermal coupling effects exist, so the only coupling effects are due to internal actions. $\B$ and $\Bc$ are square symmetric, $R_1^B=r_1^B=0$. 

\begin{figure}[h]
\center
\includegraphics[width=.32\textwidth]{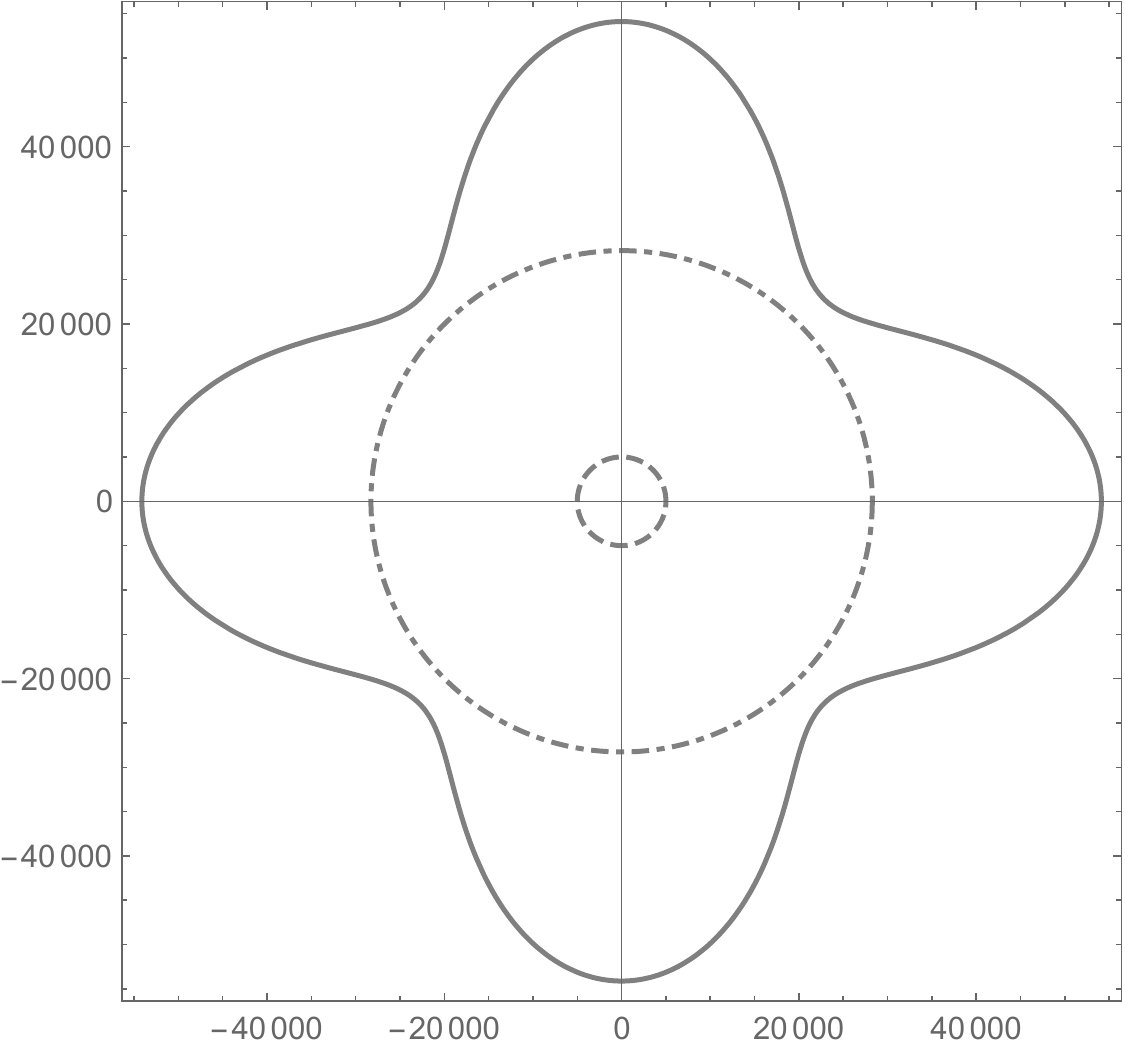}
\includegraphics[width=.32\textwidth]{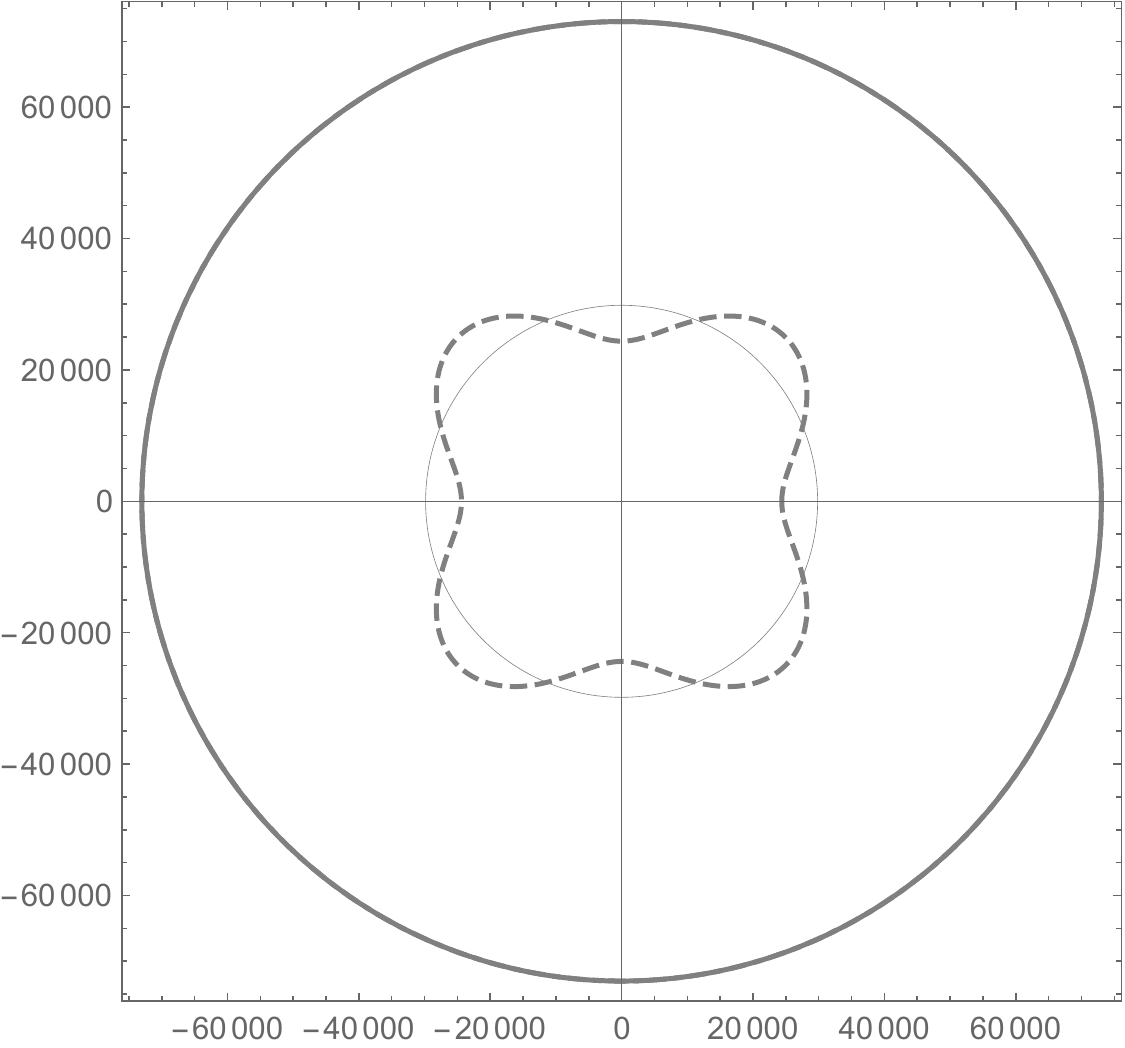}
\includegraphics[width=.326\textwidth]{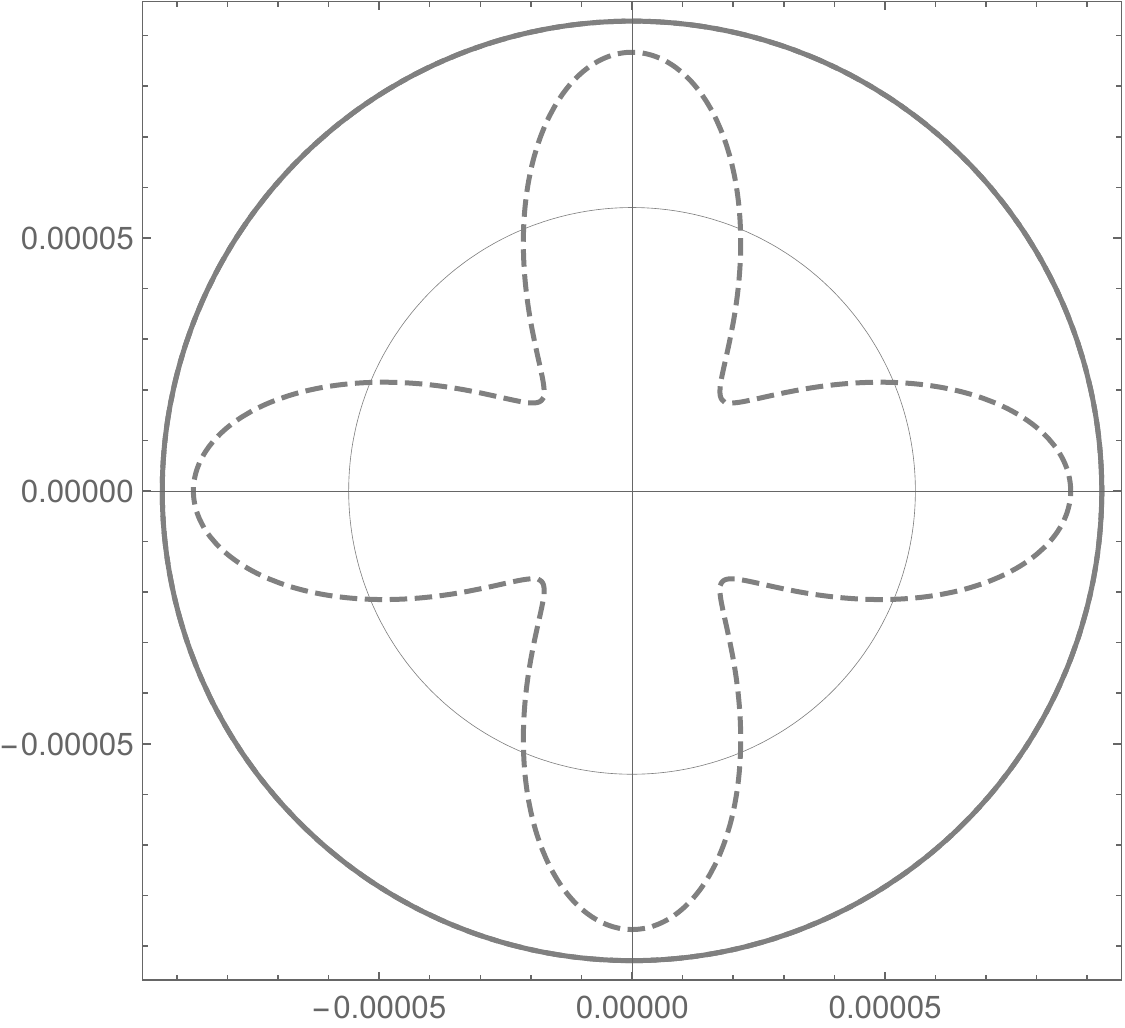}
\caption{From the left, the polar diagrams of: $\Q_{11}$, continuous line, $\alpha_1(\times10^{10})$, dashed line and $\gamma_1(\times10^6)$, dot-dashed line; $\A_{11}=\D_{11}$, continuous line, and $\B_{11}$, dashed line; $\Ac_{11}=\Dc_{11}$, continuous line, $\Bc_{11}$, dashed line.  The thin line is the {\it zero-line}: inside it, a diagram represents a negative value.}
\label{fig:4}
\end{figure}

We consider the bending of the plate due to coupling and produced by membrane forces; by eq. (\ref{eq:fundlawinversetherm})
\be
\label{eq:curvaturestrano}
\kappa_1=-\kappa_2=\frac{2}{h^2}r_0^B(N_1-N_2),\ \kappa_6=-\frac{4}{h^2}r_0^BN_6.
\ee
 In Fig. \ref{fig:4} the polar diagrams of different characteristics of the material and of the plate are presented (those not shown are isotropic).

\begin{figure}[h]
\center
\includegraphics[width=.4\textwidth]{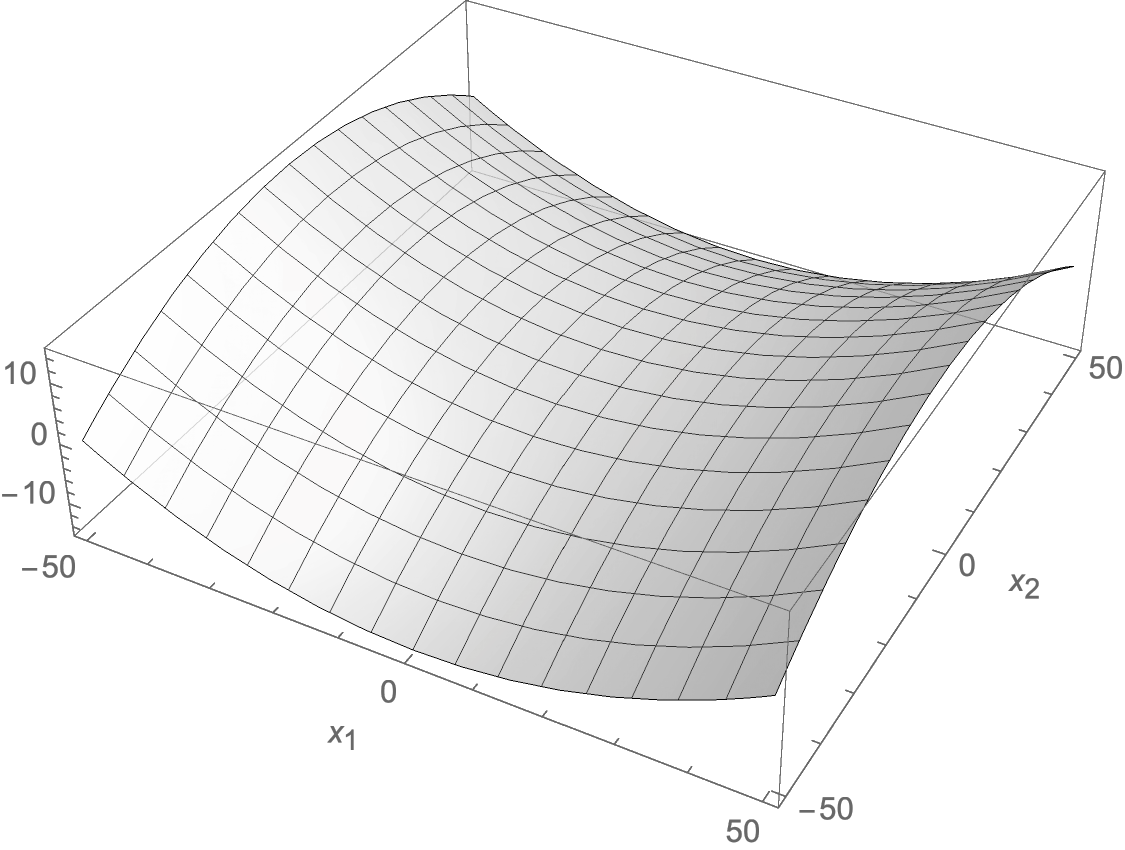}
\includegraphics[width=.4\textwidth]{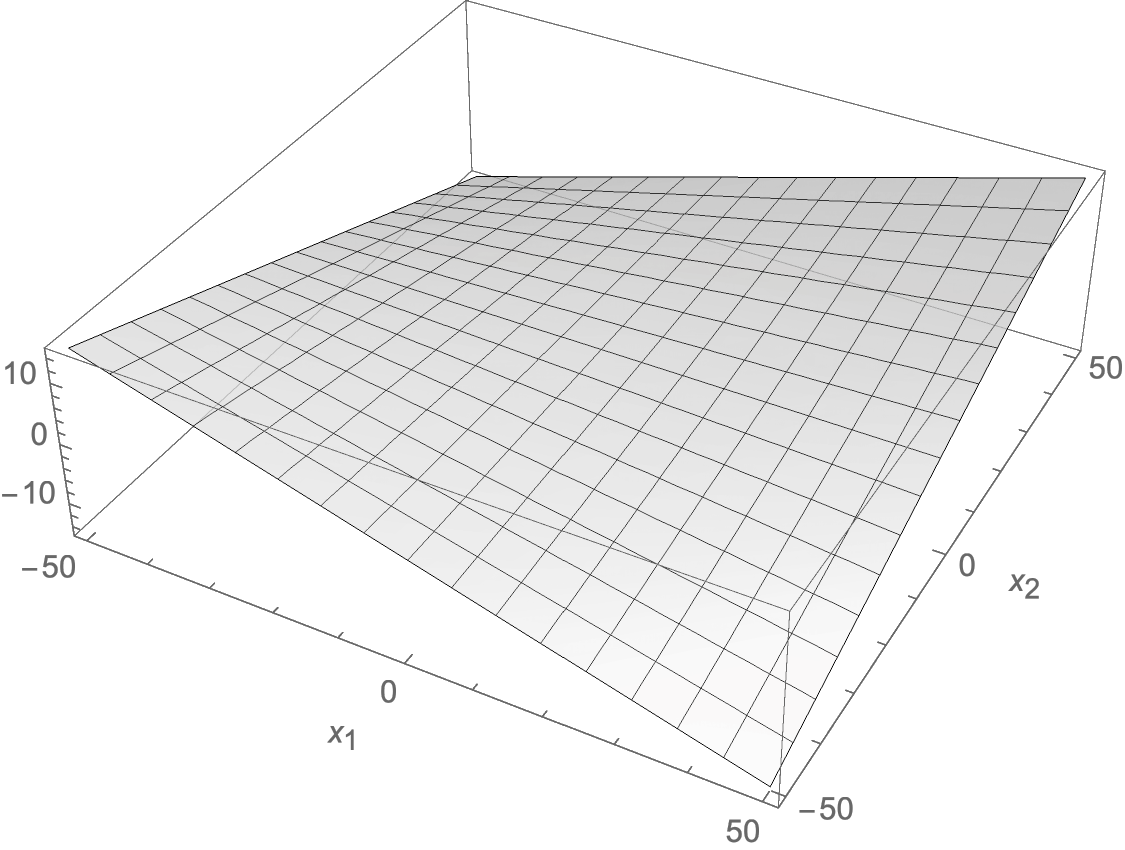}
\caption{Bending due to coupling of the $R_1^B=0$ laminate for the cases $N_1=-N_2=10$ N/mm, $N_6=0$, left, and $N_2=N_1,N_6=-10$ N/mm, right. Units are mm.}
\label{fig:5}
\end{figure}

The elastic properties of the ply are, \cite{gay14}:  $E_1=E_2=5.4\times10^4$ MPa, $G_{12}=4\times10^3$ MPa, $\nu_{12}=0.045$ and the  thickness  is 0.16 mm. The plate, whose layers have the orientations $-22.5^\circ,22.5^\circ$, has the following characteristics: dimensions $50\times50$ mm, $h=0.32$ mm, $T_0=1.49\times10^4$ MPa and $R_0^B=5.45\times10^3$ MPa.  According to eq. (\ref{eq:complianceV0}) we get $r_0^B=3.07\times10^{-5}$ MPa$^{-1}$. We consider two different cases of actions: first,  $N_1=10$ N/mm, $N_2=-N_1$ which gives, eq. (\ref{eq:curvaturestrano}), $\kappa_1=1.2\times10^{-2}$ mm$^{-1},\kappa_2=-\kappa_1,\kappa_6=0$. Then,  $N_1=N_2,N_6=-10$ N/mm, which gives, eq. (\ref{eq:curvaturestrano}), $\kappa_1=\kappa_2=0, \kappa_6=2.4\times10^{-3}$ mm$^{-1}$. These two cases are represented in Fig. \ref{fig:5}. 

\subsection{Example 3: laminate with $R_0^B=0$}
\begin{figure}[h]
\center
\includegraphics[width=.5\textwidth]{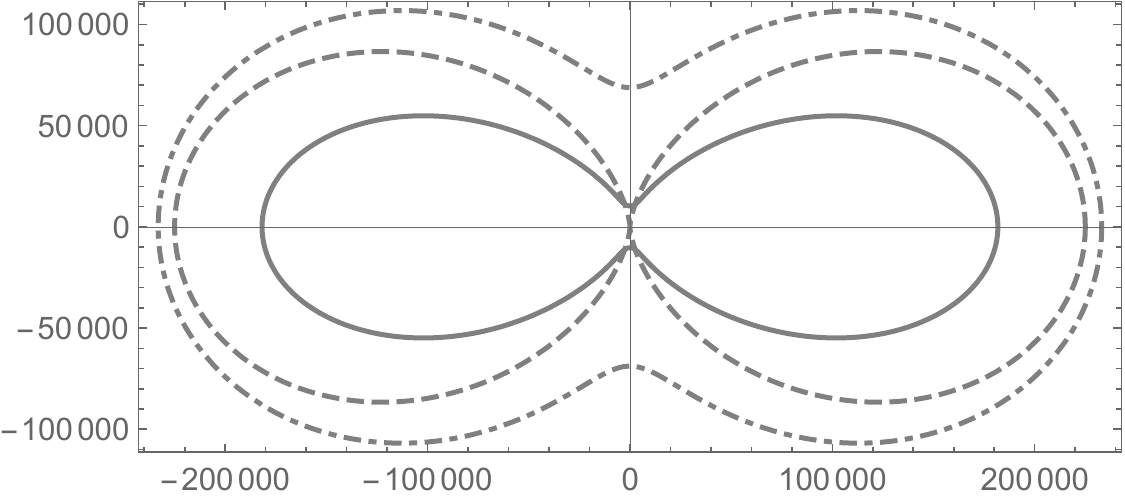}
\caption{Polar diagrams of $\Q_{11}$, continuous line, $\alpha_1(\times10^8)$, dashed line and $\gamma_1(\times10^4)$, dot-dashed line.}
\label{fig:6}
\end{figure}

This is the case of all the plates with 16 plies in Tab. \ref{tab:1}. All of them share the same $\A=\D$ and $\U=\W$, but differ for $\B$ and $\V$ and by consequence $\Bc,\u,\v_1,\v_2$ and $\w$ change for each laminate. We consider here the first sequence among the six in Tab. \ref{tab:1}. The plies are carbon-epoxy T300-5208 layers, \cite{TsaiHahn}, whose characteristics are shown in Tab. \ref{tab:4}. The characteristics of the laminate, with a thickness $h=2$ mm, are presented in Tab. \ref{tab:5}.
\begin{table}[h]
\caption{Characteristics of carbon-epoxy T300-5208 plies.}
\begin{small}
\begin{center}
\begin{tabular}{lll}
\toprule
$E_1$= 181000 MPa&$T_0$= 26880 MPa&$T_\gamma$= 15.1 MPa$^\circ$C$^{-1}$\\
$E_2$= 10300 MPa&$T_1$= 24740 MPa&$R_\gamma$= 8.21 MPa$^\circ$C$^{-1}$\\
$G_{12}$= 7170 MPa&$R_0$= 19710 MPa&$\Phi_\gamma=\ 0^\circ$\\
$\nu_{12}$= 0.28&$R_1$=21430 MPa&$h$= 0.125 mm\\
$\alpha_1=\ 2\times10^{-6}$$^\circ$C$^{-1}$&$\Phi_0=\ 0^\circ$\\
$\alpha_2=\ 2.25\times10^{-3}$$^\circ$C$^{-1}$&$\Phi_1=\ 0^\circ$\\
\bottomrule
 \end{tabular}
\end{center}
 \end{small}
\label{tab:4}
\end{table}
The polar diagrams of $\Q_{11},\alpha_1$ and $\gamma_1$ are shown in Fig. \ref{fig:6}.

\begin{table}[h]
\caption{Parameters of the 16-ply laminate with $R_0^B=0$. The  moduli not shown are null.}
\begin{small}
\begin{center}
\begin{tabular}{ll}
\toprule
\multicolumn{2}{c}{Elastic parameters}\\
\midrule
$T_0^A=T_0^D$= 26880 MPa&$T_1^A=T_1^D$= 24740 MPa\\
$R_1^B$= 3684 MPa&$\Phi_1^B=\ -90^\circ$\\
$t_0^A=t_0^D$= 9.95$\times10^{-6}$ MPa$^{-1}$&$t_1^A=t_1^D$= 2.88$\times10^{-6}$ MPa$^{-1}$\\
$r_0^A=r_0^D$= 6.49$\times10^{-6}$ MPa$^{-1}$&$\phi_0^A=\phi_0^D=0^\circ$\\
$r_1^B$= 2.37$\times10^{-6}$ MPa$^{-1}$&$\phi_1^B=\ 0^\circ$\\
\toprule
\multicolumn{2}{c}{Thermoelastic parameters}\\
\midrule
$T^ U=T^W$= 15.10 MPa $^\circ$C$^{-1}$&$R^V$= 1.41 MPa $^\circ$C$^{-1}$\\
$\Phi^V=0^\circ$&$\rho=3.83\times10^{-4}$\\
$t^u=t^w$= 1.87$\times10^{-4}$$^\circ$C$^{-1}$\\
$r^v_1$= 2.33$\times10^{-4}$ (mm $^\circ$C)$^{-1}$&$\phi^v_1=0^\circ$\\
$r^v_2$= 7.76$\times10^{-5}$ mm $^\circ$C$^{-1}$&$\phi^v_2=0^\circ$\\
\bottomrule
 \end{tabular}
  \end{center}
 \end{small}
\label{tab:5}
\end{table}
We see hence that, as foreseen,  $\B$ is $R_0$-orthotropic, like $\Bc$, $\Ac=\Dc$ are square symmetric ($r_1^A=r_1^D=0$), $\u=\w$ are isotropic and $\v_1,\v_2$ purely anisotropic ($t^v_1=t^v_2=0$). 

\begin{figure}[h]
\center
\includegraphics[width=.7\textwidth]{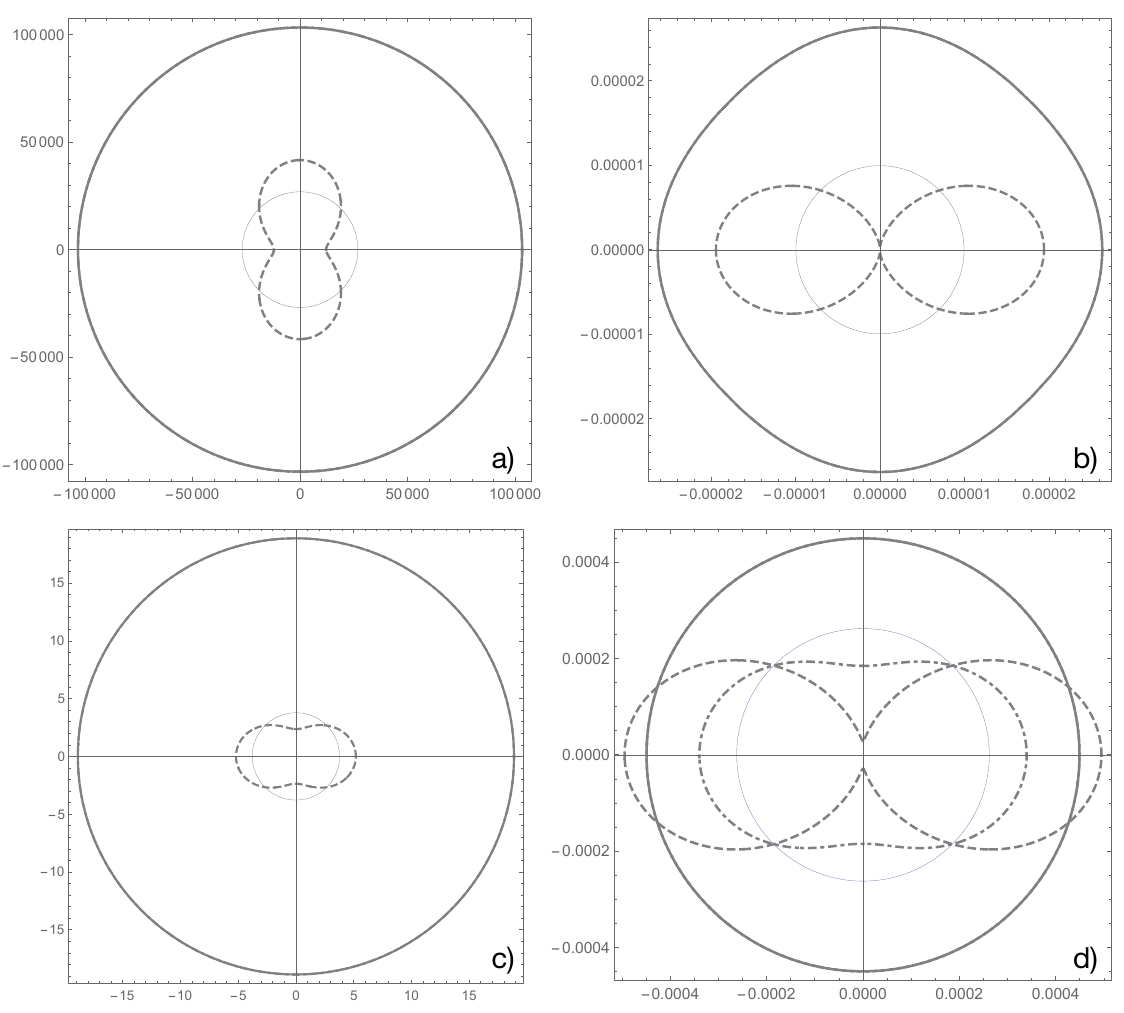}
\caption{Polar diagrams for the tensors of the 16-ply laminate with $R_0^B=0$: a) $\A_{11}=\D_{11}$, continuous line, $\B_{11}$, dashed line; b) $\Ac_{11}=\Dc_{11}$, continuous line, $\Bc_{11}$, dashed line; c) $U_{1}=W_{1}$, continuous line, $V_{1}$, dashed line; d) $u_{1}=w_{1}$, continuous line, ${(v_1)}_1$, dashed line, ${(v_2)}_1$, dot-dashed line. The thin line is the {\it zero-line}: inside it, a diagram represents a negative value.}
\label{fig:7}
\end{figure}

The polar diagrams for the various tensors of the laminate are shown in Fig. \ref{fig:7}.

\begin{figure}[h]
\center
\includegraphics[width=.32\textwidth]{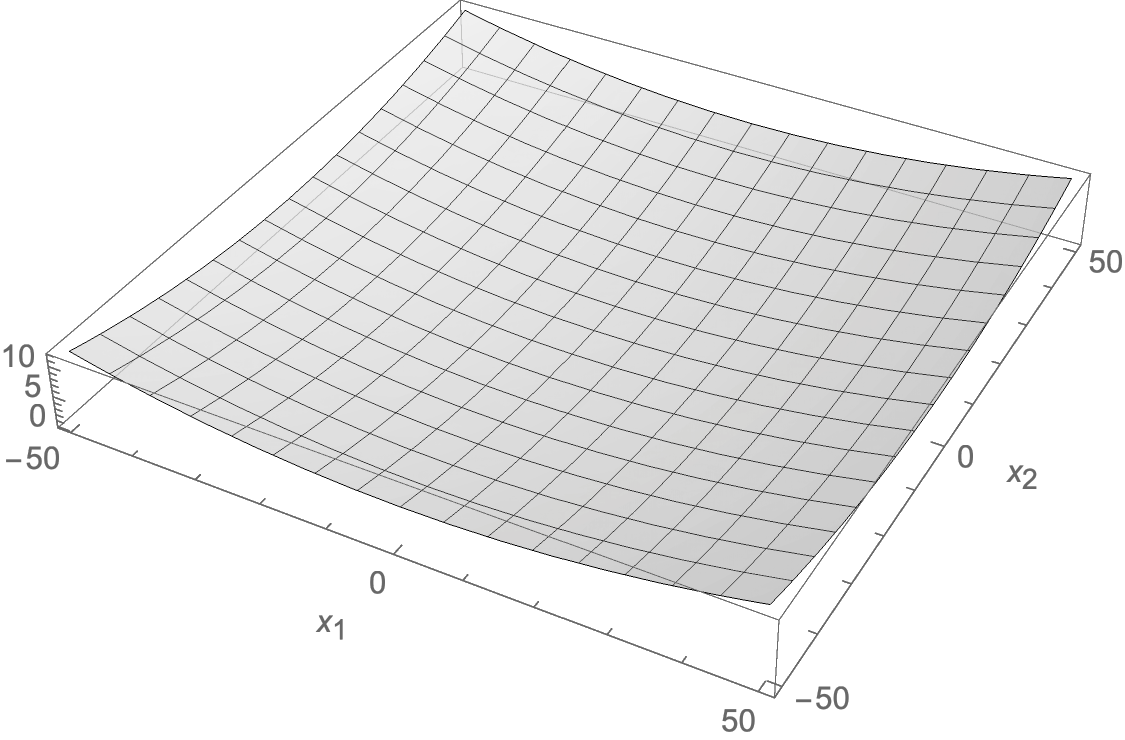}
\includegraphics[width=.32\textwidth]{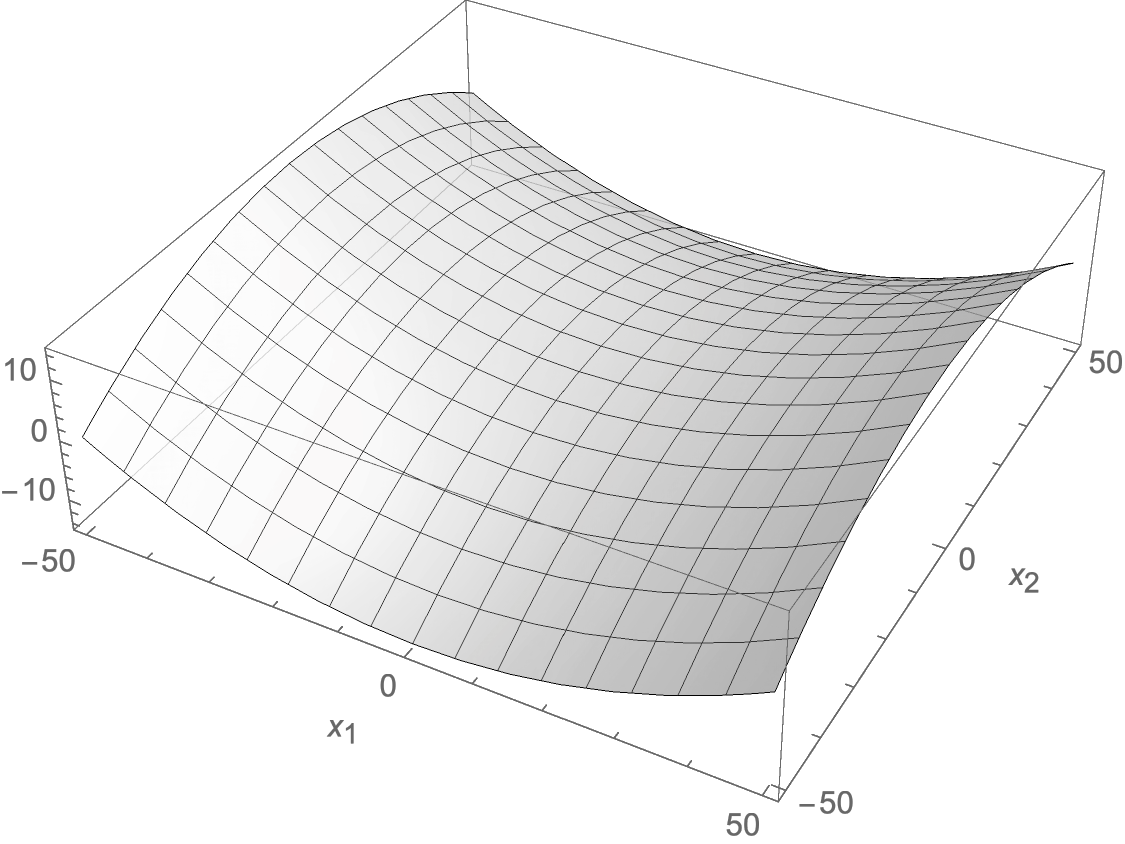}
\includegraphics[width=.32\textwidth]{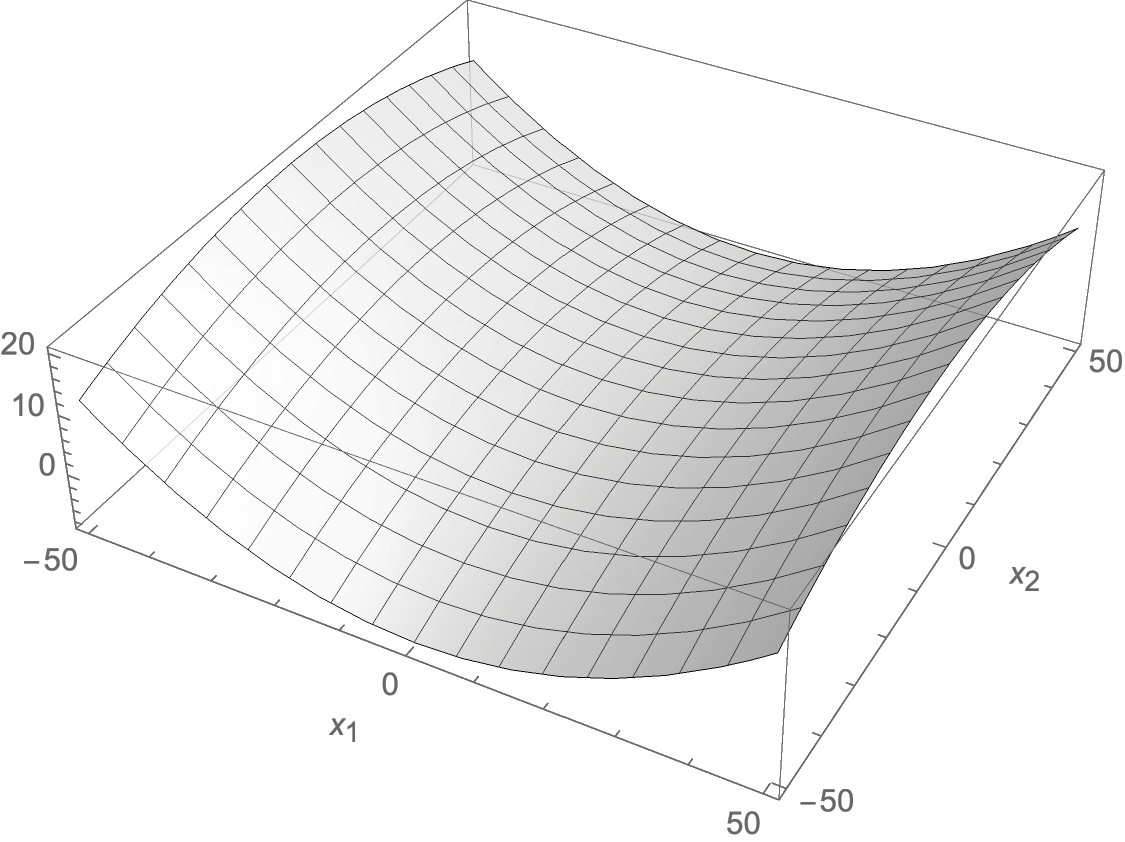}
\caption{Bending due to coupling of the 16-ply  $R_0^B=0$ laminate for the cases $N_1=-N_2=1000$ N/mm, left, $t=50^\circ$C, center, and the combined effect, right. Units are mm.}
\label{fig:8}
\end{figure}

We consider here the effects of coupling dues to membrane forces or temperature changes $t$. From eqs. (\ref{eq:fundlawinversetherm}) and (\ref{eq:casoRB0}) we get that, if $\N=(N_1,N_2,N_6)$, 
\be
\kappa_1=\frac{8}{h^2}r_1^B N_1+t\ r^v_1,\ 
\kappa_2=-\frac{8}{h^2}r_1^B N_2-t\ r^v_1,\ 
\kappa_6=0.
\ee
In Fig. \ref{fig:8} we show the deformed shape of the plate separately for the case $N_1=-N_2=1000$ N/mm and for $t=50^\circ$C, still for a square plate  with a side of 100 mm, and the combination of the two effects. 

\subsection{Example 4: laminate with anisotropic $\B$}
\begin{figure}[h]
\center
\includegraphics[width=.7\textwidth]{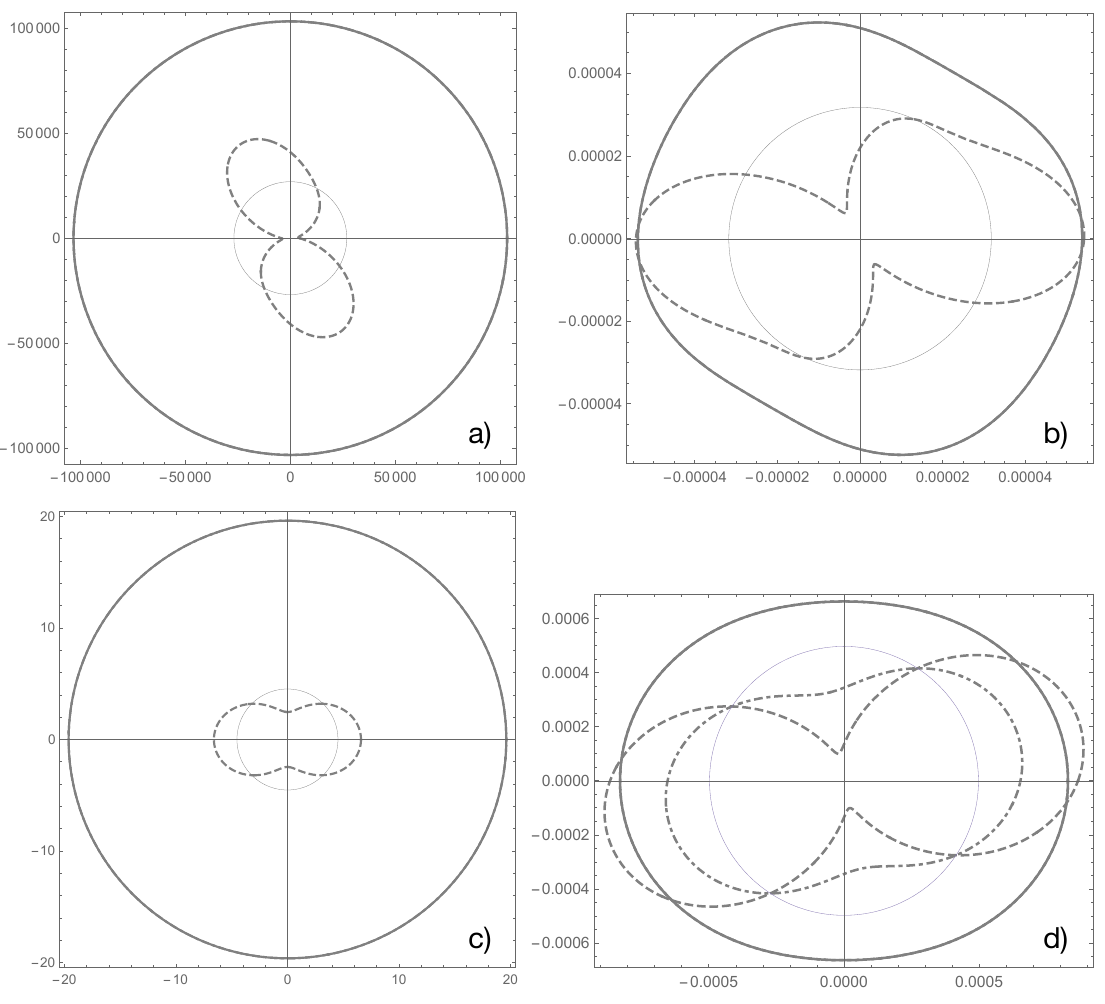}
\caption{Polar diagrams for the tensors of the 18-ply laminate: a) $\A_{11}=\D_{11}$, continuous line, $\B_{11}$, dashed line; b) $\Ac_{11}=\Dc_{11}$, continuous line, $\Bc_{11}$, dashed line; c) $U_{1}=W_{1}$, continuous line, $V_{1}$, dashed line; d) $u_{1}=w_{1}$, continuous line, ${(v_1)}_1$, dashed line, ${(v_2)}_1$, dot-dashed line. The thin line is the {\it zero-line}: inside it, a diagram represents a negative value.}
\label{fig:9}
\end{figure}

The last example is that of the first 18-ply laminate in Tab. \ref{tab:1}. Like before, all the stacks have the same $\A=\D$ and $\U=\W$, but differ for $\B$ and $\V$ and by consequence $\Bc,\u,\v_1,\v_2$ and $\w$ change for each laminate.  The material composing the plate is still carbon-epoxy T300-5208, Tab. \ref{tab:4} and Fig. \ref{fig:6}. The value of $\rho$, that depends on the material, does not change with respect to the previous example. Moreover, because of isotropy, $\A=\D$ and $\U=\V$ are the same of the previous case, as they are determined by the isotropic part of the material, which is the same for the two examples. The parameters of the plate, whose thickness is $h=2.25$ mm, are presented in Tab. \ref{tab:6}. The polar diagrams for the plate are shown in Fig. \ref{fig:9}.

\begin{table}[h]
\caption{Parameters of the 18-ply laminate. The  moduli not shown are null.}
\begin{small}
\begin{center}
\begin{tabular}{lll}
\toprule
\multicolumn{3}{c}{Elastic parameters}\\
\midrule
$T_0^A=T_0^D$= 26880 MPa&$T_1^A=T_1^D$= 24740 MPa\\
$R_0^B$= 4956 MPa&$R_1^B$= 3684 MPa\\
$\Phi_0^B=\ 36.972^\circ$&$\Phi_1^B=\ -73.945^\circ$\\
$t_0^A=t_0^D$= 1.27$\times10^{-5}$ MPa$^{-1}$&$t_1^A=t_1^D$= 3.57$\times10^{-6}$ MPa$^{-1}$\\
$r_0^A=r_0^D$= 1.92$\times10^{-6}$ MPa$^{-1}$&$r_1^A=r_1^D$= 8.86$\times10^{-7}$ MPa$^{-1}$\\
$\phi_0^A=\phi_0^D=16.556^\circ$&$\phi_1^A=\phi_1^D=-33.396^\circ$\\
$t_0^B=\ -3.19\times10^{-7}$ MPa$^{-1}$&$t_1^B=\ -8.96\times10^{-8}$ MPa$^{-1}$\\
$r_0^B$= 8.09$\times10^{-6}$ MPa$^{-1}$&$r_1^B$= 4.78$\times10^{-6}$ MPa$^{-1}$\\
$\phi_0^B=\ -8.112^\circ$&$\phi_1^B=\ 16.187^\circ$\\
\toprule
\multicolumn{3}{c}{Thermoelastic parameters}\\
\midrule
$T^ U=T^W$= 15.10 MPa $^\circ$C$^{-1}$&$R^V$= 2.06 MPa $^\circ$C$^{-1}$&$\Phi^V=0^\circ$\\
$t^u=t^w$= 2.49$\times10^{-4}$$^\circ$C$^{-1}$&$r^u=r^w$= 8.25$\times10^{-5}$$^\circ$C$^{-1}$&$\phi^u=\ -27.219^\circ$\\
$t^v_1$= 2.88$\times10^{-6}$ (mm $^\circ$C)$^{-1}$&$r^v_1$= 3.97$\times10^{-4}$ (mm $^\circ$C)$^{-1}$&$\phi^v_1=11.616^\circ$\\
$t^v_2$= 1.22$\times10^{-6}$ mm $^\circ$C$^{-1}$&$r^v_2$= 1.68$\times10^{-4}$ mm $^\circ$C$^{-1}$&$\phi^v_2=11.616^\circ$\\
\bottomrule
 \end{tabular}
  \end{center}
 \end{small}
\label{tab:6}
\end{table}
We remark that now $\B$ has the two anisotropic phases: $R_0^B\neq0,R_1^B\neq0$. As a consequence, $\Ac=\Dc$ are completely anisotropic, not even orthotropic; $\u=\w$ are orthotropic as well as $\v_1$ and $\v_2$. The orientation of the tensors, i.e. the angles $\Phi_1^B,\phi_1^A=\phi_1^D,\phi_1^B,\phi^u=\phi^w,\phi_1^V=\phi_2^v$ depend on the stacking sequence. The response of the plate to applied actions and/or temperature changes is much more complicate, as ruled by completely anisotropic tensors.

\section{Conclusion}
We have seen how an apparently simple class of laminated plates, those having isotropic extension and in bending stiffnesses, can indeed have a rather complicate mechanical response when the plate is coupled. In such a situation, even define the material symmetries of the plate becomes questionable, because the compliance behaviors in general differ from the corresponding stiffness ones. Moreover, the numerous differences between the case of hybrid and identical plies laminates have been put in evidence. 

This type of laminates, because of their so peculiar and articulated response, can constitute an interesting possibility for the realization of mechanical devices where interaction between forces or  temperature changes and deformations are to be designed to some purpose. Research is going on in this direction.

\bibliographystyle{vancouver} 
\bibliography{Biblio}   

\end{document}